\documentclass[letterpaper]{mn2e}
\usepackage{times}
\usepackage{graphicx}
\usepackage{soul}       %%%%% REMOVE IN SUBMITTED VERSION!!!
\usepackage{amsmath}
\usepackage{txfonts}
\def\Msolar{\hbox{${\rm M}_\odot$}}

\begin{document} 

\title[Extinction law in the Tarantula nebula]{Hubble Tarantula Treasury
Project. IV. The extinction law\thanks{Based on observations with the 
NASA/ESA {\it Hubble Space Telescope}, obtained at the Space Telescope
Science Institute, which is operated by AURA, Inc., under NASA contract
NAS5-26555.}}

\author[Guido De Marchi et al.]
{Guido~De~Marchi$^1$, Nino~Panagia$^{2,3,4}$, Elena Sabbi$^2$, Daniel
Lennon$^5$, Jay Anderson$^2$,
\newauthor 
Roeland van der Marel$^2$, Michele~Cignoni$^2$, Eva~K.~Grebel$^6$, 
S{\o}ren Larsen$^7$, 
\newauthor
Dennis Zaritsky$^8$, Peter Zeidler$^6$, Dimitrios Gouliermis$^9$, 
Alessandra Aloisi$^2$\\
$^1$European Space Research and Technology Centre, Keplerlaan 1, 2200 AG
Noordwijk, The Netherlands, gdemarchi@esa.int \\
$^2$Space Telescope Science Institute, 3700 San Martin Drive, Baltimore, MD
21218, USA\\
$^3$INAF--NA, Osservatorio Astronomico di Capodimonte, Salita Moiariello
16, 80131 Naples, Italy\\
$^4$Supernova Ltd, OYV \#131, Northsound Rd., Virgin Gorda VG1150,
Virgin Islands, UK\\
$^5$European Space Astronomy Centre, Apdo. de Correo 78, 28691
Villanueva de la Ca\~nada, Madrid, Spain\\
$^6$Astronomisches Rechen-Institut, 
Zentrum f\"ur Astronomie der Universit\"at Heidelberg, M\"onchhofstr.\ 
12--14, 69120 Heidelberg, Germany\\
$^7$Department of Astrophysics, Radboud University, P.O. Box 9010, 6500 GL 
Nijmegen, The Netherlands\\
$^8$Steward Observatory, University of Arizona, 933 N. Cherry Ave, Tucson, 
AZ 85721-0065, USA\\
$^9$ Institut f\"ur Theoretische Astrophysik, 
Zentrum f\"ur Astronomie der Universit\"at Heidelberg,
Albert-Ueberle-Str. 2, 69120 Heidelberg, Germany}

\date{Received 1.9.2015; Accepted 27.10.2015}
\pagerange{\pageref{firstpage}--\pageref{lastpage}} \pubyear{2015}

\maketitle

\begin{abstract}

We report on the study of interstellar extinction across the Tarantula
nebula (30\,Doradus), in the Large Magellanic Cloud, using observations
from the Hubble Tarantula Treasury Project in the $0.3-1.6\,\muup$m
range. The considerable and patchy extinction inside the nebula causes
about $3\,500$ red clump stars to be scattered along the reddening
vector in the colour--magnitude diagrams, thereby allowing an accurate
determination of the reddening slope in all bands. The measured {  slope
of the reddening vector is remarkably steeper in all bands} than in the
the Galactic diffuse interstellar medium. At optical wavelengths, the
larger ratio of total-to-selective extinction, namely $R_V=4.5 \pm 0.2$,
implies the presence of a grey component in the extinction law, due to a
larger fraction of large grains. The extra large grains are most likely
ices from supernova ejecta and will significantly alter the extinction
properties of the region until they sublimate in $50-100$\,Myr. We
discuss the implications of this extinction law for the Tarantula nebula
and in general for regions of massive star formation in galaxies. Our
results suggest that fluxes of strongly star forming regions are likely
to be underestimated by a factor of about 2 in the optical. 

\end{abstract}


\section{Introduction} 

The Hubble Tarantula Treasury Project (HTTP) is a photometric survey at 
high spatial resolution of the Tarantula nebula (30\,Dor), from near
ultraviolet (NUV) to near infrared (NIR) wavelengths (Sabbi et al.
2013). Its purpose is to reconstruct the region's star-formation history
in space and time on a parsec scale over a total extent of $\sim 240
\times 190$\,pc$^2$. the ultimate goal is to establish the strength,
duration, and spatial scale of the star-formation episodes and their
possible mutual relationships. An initial study limited to the central
NGC\,2070 cluster (Cignoni et al. 2015) confirms that over the past
$\sim 20$\,Myr the cluster experienced a prolonged activity of star
formation  (e.g., Walborn \& Blades 1997; {  Walborn et al. 1999}),
with several episodes (De Marchi et al. 2011a), culminating in a peak
$\sim 1 - 3$\,Myr ago.

Besides high-quality photometry (Sabbi et al. 2015), these studies rely
on our ability to securely measure the intrinsic physical properties of
stars, i.e. their true colours and luminosities, since these are crucial
to extract reliable masses, ages and other physical parameters to track
the star-formation process. Knowledge of the properties and amount of
the interstellar extinction is thus of paramount importance, and equally
fundamental is knowing how to apply this information to correct the
photometry of individual stars. This is particularly crucial in an
environment such as the Tarantula nebula, due to its complex structure
and to the presence of a considerable amount of atomic and molecular gas
(e.g., Indebetouw et al. 2013; Yeh et al. 2015) and dust (e.g., Meixner
et al. 2013), resulting in a patchy and uneven level of extinction
across the nebula. 

High resolution {\em Hubble Space Telescope} (HST) studies of the
massive stellar Tarantula clusters NGC\,2070 and Hodge\,301 have long
shown a wide spread of extinction values (e.g., Hunter et al. 1995a;
Grebel \& Chu 2000; De Marchi et al. 2011a), particularly in the central
NGC\,2070 cluster, where $A_V$ varies by more than 3\,mag over regions
of $\sim 40$\,pc across, as shown by Ma\'{\i}z Apell\'aniz et al. (2014)
and De Marchi \& Panagia (2014). More importantly, both Ma\'{\i}z
Apell\'aniz et al. (2014) and De Marchi \& Panagia (2014) have
independently shown from spectroscopy and photometry that the extinction
law for the central NGC\,2070 cluster is very different from that
typical of the diffuse Galactic interstellar medium (ISM), with a $\sim
50\,\%$ higher ratio of total-to-selective extinction, namely 
$R_V=A_V/E(B-V)=4.5$ instead of $3.1$. 

{\rm This finding is particularly intriguing because the Tarantula
nebula is routinely considered an ideal test case (the ``Starburst
Rosetta'', Walborn 1991) for regions of strong star formation at greater
distances, where observations cannot reveal individual objects and one
must rely on integrated properties. Therefore, understanding whether
these apparently anomalous extinction properties are just peculiar to
the central NGC\,2070 cluster or are a common feature throughout the
Tarantula complex is fundamental for our understanding and
interpretation of the integrated star formation diagnostics and of the
chemical evolution in more distant galaxies. }

In this paper, we extend the work of De Marchi \& Panagia (2014) to the
entire Tarantula nebula. The traditional approach to determine the
extinction properties is the ``pair method'', whereby the spectrum of a 
reddened star is compared with that of a reference, un-extinguished
object of the same spectral type (e.g., Johnson 1968; Massa, Savage \&
Fitzpatrick 1983; Cardelli, Sembach \& Mathis 1992). This requires high
quality spectra, extending from the NUV to the NIR, that are difficult
to obtain in the crowded 30\,Dor regions and are necessarily limited to
the brightest and hence most massive stars (e.g., Fitzpatrick \& Savage
1984; Gordon et al. 2003; Ma\'{\i}z Apell\'aniz et al. 2014). Following
this method results in a sparse  coverage of the Tarantula nebula,
preferentially limited to the areas of more recent star formation.
Conversely, by making use of multi-band photometry of red giant stars in
the red clump (RC) phase (e.g., Paczynski \& Stanek 1998; Cole 1998;
Girardi et al. 1998; { Gao et al. 2009; Wang et al. 2013}), the
method developed by De Marchi \& Panagia (2014; see also De Marchi,
Panagia \& Girardi 2014) allows us to obtain a rich and uniform coverage
spread over several thousand lines of sight  in the Tarantula region,
resulting in a self-consistent absolute extinction curve of high
statistical significance over the entire field and wavelength range of
the observations.

The structure of the paper is as follows. In Section 2 we describe the
HTTP observations relevant for this study. Section 3 is devoted to the
identification of RC stars through an innovative use of unsharp-masking
techniques. In Section 4 we derive the absolute extinction towards RC 
stars and the corresponding extinction law. In Section 5 we present the 
reddening distribution in this field and discuss how this information
should be used to correct the photometry of individual objects. A
summary and our conclusions follow in Section 6.

\section{Observations}

The observations are part of the HTTP survey, described in detail in
Sabbi et al. (2013, 2015). They cover a region of $16^\prime \times
13^\prime$ including the 30\,Dor nebula, corresponding to $\sim 240
\times 190$\,pc$^2$ at the distance of the Large Magellanic Cloud (LMC).
{  Throughout this paper we will adopt a distance modulus $(m-M)_0=18.55
\pm 0.05$ as obtained by Panagia et al. (1991; see also Panagia 2005)
for SN\,1987A, located in the vicinity of the Tarantula nebula.} The
observations were obtained with the {\em Advanced Camera for Surveys}
(ACS) and {\em Wide Field Camera 3} (WFC3) instruments on board the {\em
HST} in a set of broad and narrow bands over the range $0.27 -
1.6$\,$\muup$m (respectively F275W, F336W, F555W, F658N, F775W, F110W,
and F160W). The photometric reduction and the corresponding catalogue
are presented in Sabbi et al. (2015). That paper also illustrates how
the two cameras were used to cover the entire field and it provides a
detailed list of the exposure times reached in each field. The latter
typically amount to 1\,164\,s in the F275W band, 1\,402\,s in F336W,
2\,270\,s in F555W, 2\,220\,s in F658N, 2\,329\,s or 2\,639\,s in F775W,
1\,298\,s in F110W, and 1\,598\,s in F160W   

\begin{figure}
\centering
%\resizebox{\hsize}{!}{\includegraphics[width=16cm]{httpfig42c.pdf}}
 \resizebox{\hsize}{!}{\includegraphics[width=16cm]{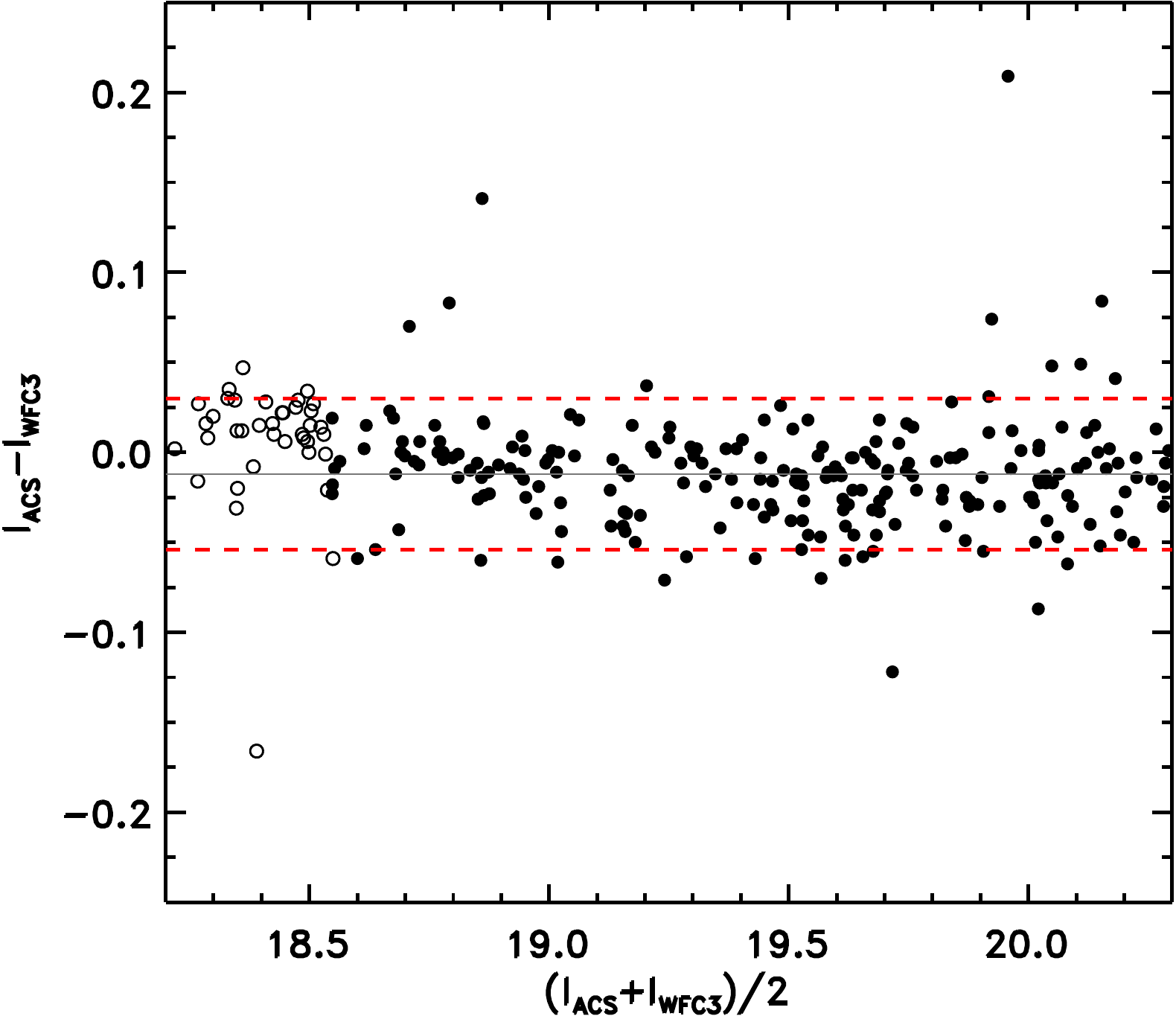}}
\caption{Comparison between the $m_{\rm 775}$ magnitudes measured for
the same stars with the ACS and WFC3. The mean magnitude difference
(solid line) is $0.006$\,mag, or less than half the typical photometric
uncertainty of $0.014$\,mag ($1 \sigma$). The dashed lines mark the $\pm
3\,\sigma$ band.}
\label{fig1}
\end{figure}

Observations in the F775W band were taken with both the ACS and WFC3,
covering adjacent regions (see Sabbi et al. 2015 for details). Both
cameras feature a filter with that name, but although rather similar 
their overall response in those bands is not quite the same. In
Figure\,\ref{fig1}, we show the differences between the F775W magnitudes
of objects in a strip of $\sim 16^\prime \times 1^\prime$  that was
observed in this band with both cameras. The selected magnitude range,
$18 \la m_{\rm 775} \la 20$, is relevant for the RC stars discussed in
this work. The thin solid line shows the mean magnitude difference
between the two bands, corresponding to $0.006$\,mag. The difference is
smaller than the typical photometric uncertainty for these objects,
namely $\sim 0.014$\,mag (the dashed lines mark the corresponding $\pm
3\,\sigma$ band). The root mean square deviation with respect to the
mean is $0.038$\,mag, whereas the same root mean square deviation with
respect to zero is just a millimagnitude larger, or $0.039$\,mag, making
a correction not necessary. 

{\rm Close inspection of the trend seen in the figure might suggest that
there is a small colour term, since for this population the magnitude
correlates directly with the colour. However, most of the stars in the
range $\sim 18.2 - 18.6$ have uncertainties estimated by Sabbi et al.
(2015) to be $0.1$\,mag due to saturation (open circles in
Figure\,\ref{fig1}). For this reason, the apparent deviation is not
significant. Indeed, as we will show in Section\,4, there is no
detectable systematic difference between the slopes of the reddening
vectors in the northern and southern portions of the field, covered
respectively by the WFC\,3 and ACS. Therefore, in the context of this
work it is not  necessary to apply a colour-term correction to the
photometry and in the following we will not distinguish between the two
F775W bands. }

Note that in Figure\,\ref{fig1} there are a few stars with differences
exceeding $3\,\sigma$ (dashed lines). {\rm Even though their number is
not statistically significant, these objects} {  could be variable stars
in the field}, and the magnitude difference may originate because the
ACS and WFC3 observations were not taken simultaneously, {  or they
might be blends or stars with nearby neighbours in projection.} 

%% At the bright end ($18.0 - 18.2$) it appears that the ACS
%% magnitude is systematically fainter, by $\sim 0.02$\,mag, but
%% this has no effect on reddened RC stars, which are all fainter
%% than $m_{775} = 18.2$ (see Table 1).  

\begin{figure}
\centering
%\resizebox{\hsize}{!}{\includegraphics[width=16cm]{httpfig1a.pap.pdf}}
 \resizebox{\hsize}{!}{\includegraphics[width=16cm]{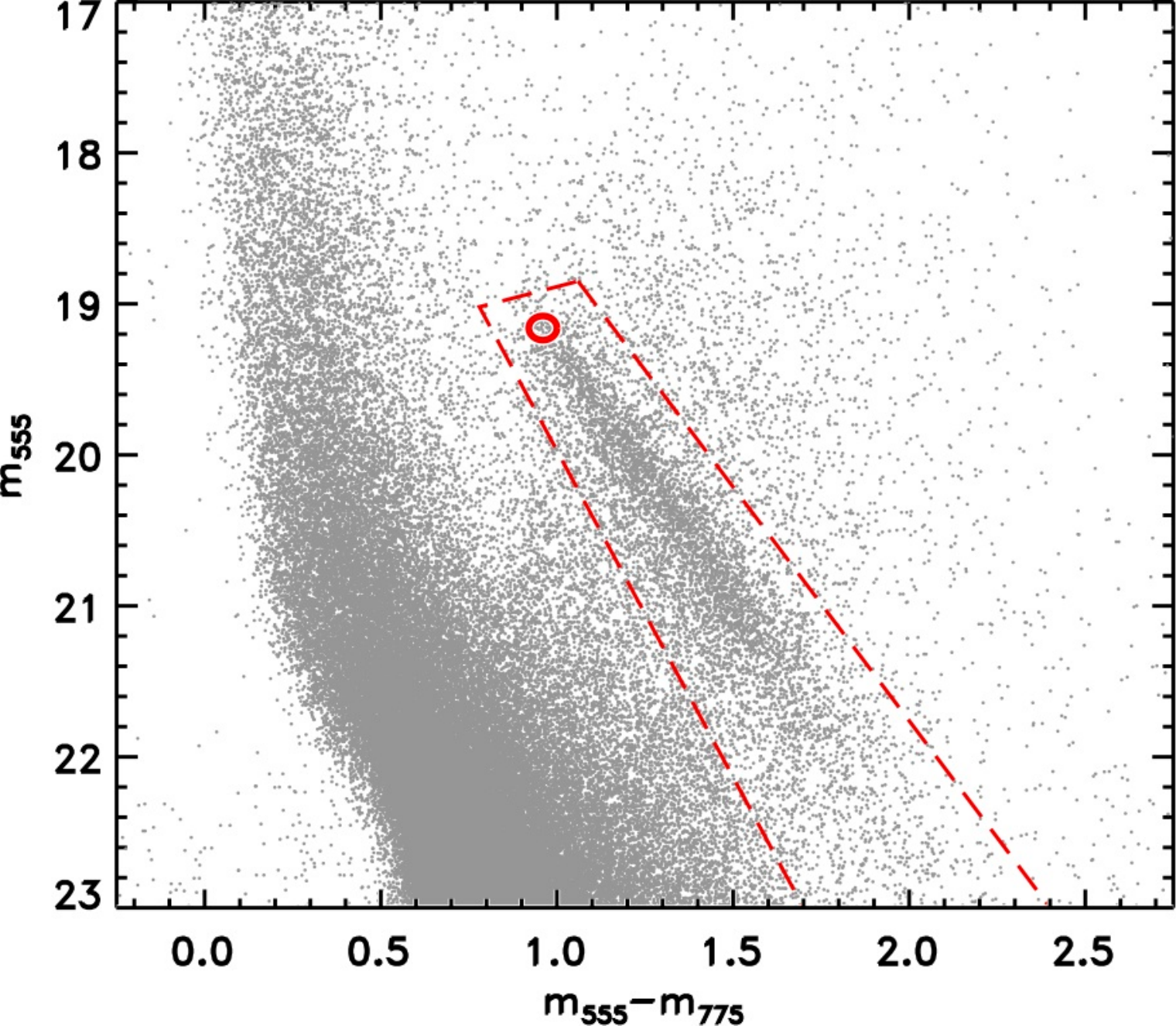}}
\caption{CMD of the entire field. The rather broad UMS
suggests the presence of considerable and uneven level of extinction in
the field. This is confirmed by the prominent RC, extending by several
magnitudes from its expected location (red circle). }
\label{fig2}
\end{figure}

Selecting from the Sabbi et al. (2015) catalogue all the stars with
combined  photometric uncertainty  $\le 0.1$\,mag in the F555W and
F775W bands ($\sim 755\,000$ objects), we obtain the colour--magnitude 
diagram (CMD) shown in Figure\,\ref{fig2}. Following Romaniello (1998),
the combined uncertainty $\delta_2$ is defined as:

\begin{equation}
\delta_2 = \sqrt{\frac{\delta^2_{555} + \delta^2_{775}}{2}}
\label{eq1}
\end{equation}

\noindent 
where $\delta_{555}$ and $\delta_{775}$ are the uncertainties in each 
individual band.\footnote{The definition given by Equation\,\ref{eq1} 
can be generalised for any combination of bands.} 
  
Besides a rather broad upper main sequence (UMS), several times wider
than the photometric uncertainty $\delta_2 < 0.1$\,mag, the most
prominent feature of Figure\,\ref{fig2} is an elongated stellar sequence
almost parallel to the main sequence (MS) itself but well separated from
it. To help characterise its nature, we  show as a red circle the
location of the ``nominal RC'', defined as the theoretical RC of stars
of the lowest metallicity applicable to this field and for ages in the
range $1.4 - 3.0$\,Gyr. De Marchi, Panagia \& Girardi (2014) have shown
that a metallicity  $Z = 0.004$ is appropriate  for the old stars ($>
1$\,Gyr) in 30\,Dor. 

\begin{table}
\centering 
\caption{Apparent magnitudes $m_{\rm RC}$ of the RC and corresponding 
$1\,\sigma$ spread in all bands, already including the effects of the
distance and extinction by intervening MW dust in the foreground.}
\begin{tabular}{llcc} 
\hline
Band &  $m_{\rm RC}$ & $\sigma$\\
\hline 
F275W & $22.20$ & $0.12$\\
F336W & $20.34$ & $0.12$\\
F438W & $19.98$ & $0.10$\\
F555W & $19.16$ & $0.08$\\
F775W & $18.21$ & $0.08$\\ 
F110W & $17.72$ & $0.10$\\
F160W & $17.15$ & $0.10$\\
\hline
\end{tabular}
\label{tab1}
\end{table}

The apparent magnitudes of the nominal RC and the $1\,\sigma$ spread
around them are listed in Table\,\ref{tab1} for the bands relevant to
this study (derived from De Marchi et al. 2014; De Marchi \& Panagia
2014). These magnitudes take account of the distance modulus $(m - M)_0
= 18.55$ (Panagia et al. 1991; Panagia 2005; Walborn \& Blades 1997) and
already include the contribution of the foreground Milky Way (MW)
absorption along the line of sight. Fitzpatrick \& Savage (1984)
estimated the latter to be $E(B-V) = 0.07$ or $A_V = 0.22$ and these are
the values that we will assume throughout this work.

The excellent agreement between the position of the nominal RC and the
head of the elongated sequence confirms that the latter is indeed made
up of RC stars, which are displaced in the CMD along the direction of
the reddening vector by the considerable and uneven levels of extinction
present in this field. {  The extended RC has been used in the past to
study the reddening distribution and to derive reddening maps in the
Magellanic clouds, having assumed an extinction law (e.g. Zaritsky 1999;
Haschke, Grebel \& Duffau 2011; Tatton et al. 2013). More recently,} De
Marchi et al. (2014) and De Marchi \& Panagia (2014) have shown how to
use the extended RC feature in CMDs like that of Figure\,\ref{fig2} to
derive the extinction law and the extinction to the individual objects
in the field. The procedure includes three main steps, namely, {\em i)}
the identification of the candidate RC stars, {\em ii)} the removal of
possible outliers, and {\em iii)} the determination of the slope of the
reddening vector in each set of bands. In general, in order to identify
candidate RC stars one needs to know from theory the intrinsic colour
and magnitude of the nominal RC, for the metallicity and distance of the
population under study (see e.g. Table\,\ref{tab1}). However, when  the
RC population consists of several thousands stars, like in the present
case, we can follow a fully empirical approach and determine the
location of the un-extinguished RC in the CMD using on it the image
sharpening technique known as ``unsharp masking.''

\section{Unsharp masking}

The technique of unsharp-masking photographic images was originally 
presented by Spiegler \& Juris (1931) and later discussed by Yule
(1944). It consists in making an image sharper by overlapping the image
itself and an inverted blurred version of it. The blurred image is an
out of focus version of the original and has to be subtracted from it
(hence it is printed in negative if the original is in positive, or
vice-versa). The blurred reversed duplicate is called a mask. Combining
the mask with the original image reduces considerably the intensity of
the low frequency features, but does not affect the high frequency
contrast. Therefore, the image appears sharper because of the
enhancement in the high frequency features. 

\begin{figure}
\centering
%\resizebox{\hsize}{!}{\includegraphics[width=16cm]{httpfig49.pdf}}
 \resizebox{\hsize}{!}{\includegraphics[width=16cm]{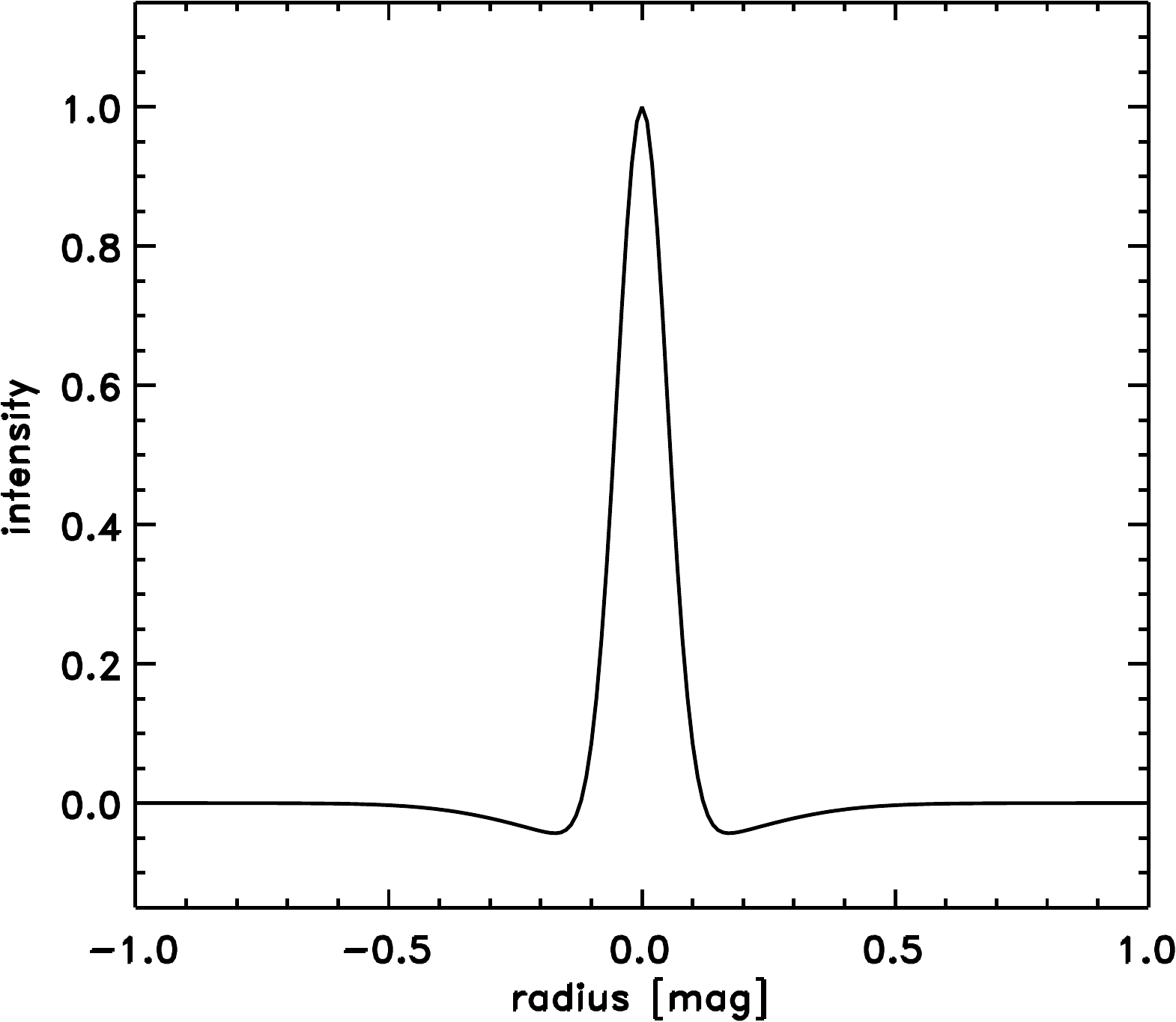}}
\caption{Profile of the kernel used for unsharp-masking the CMD of
Figure\,\ref{fig2}. }
\label{fig3}
\end{figure}

Photographic masking as a technique was regularly used in the graphic
arts industry. Its principle is today routinely applied in the 
unsharp-masking tools of modern digital-imaging software packages, such
as {\em GIMP} (Kylander \& Kylander 1999). The software subtracts from
the original a Gaussian-blurred copy of it. Conceptually, digital
unsharp-masking is equivalent to the linear operation of convolving the
original with a kernel that is the Dirac delta minus a Gaussian blur
kernel. 

In this work we use unsharp masking in an innovative way, by applying it
to the CMDs, rather than to the digital images themselves. First the
CMD is converted to an image, i.e. a two-dimensional array similar to a
Hess diagram (Hess 1924). The points in the CMD are mapped onto an array
with a sampling of $0.01$\,mag in colour and magnitude and the array is
then convolved with a narrow Gaussian beam. The convolution is meant to
assign to each CMD point the proper resolution, including uncertainties
on the photometry and on the nature of the objects, as if it were the
resolution element of an image. With a typical photometric uncertainty
of $\sim 0.025$\,mag for the stars in the CMD of Figure\,\ref{fig2}, we
have used $\sigma=0.05$\,mag, or twice the photometric uncertainty. The
second step is the creation of the mask, by convolving the resulting CMD
image with a wider Gaussian beam. We have experimented with several
values and selected $\sigma=0.2$\,mag, although values in the range
$0.15 < \sigma < 0.30$\,mag would give comparably good results. The mask
is then subtracted from the CMD image. Note that, analytically, these
operations are equivalent to convolving the CMD with a kernel. Instead
of a Dirac delta minus a Gaussian beam, our kernel is the difference
between two Gaussian beams with different $\sigma$. A radial section of
the resulting kernel is shown in Figure\,\ref{fig3}.

\begin{figure}
\centering
%\resizebox{\hsize}{!}{\includegraphics[width=16cm]{httpfig1b.pap.pdf}}
 \resizebox{\hsize}{!}{\includegraphics[width=16cm]{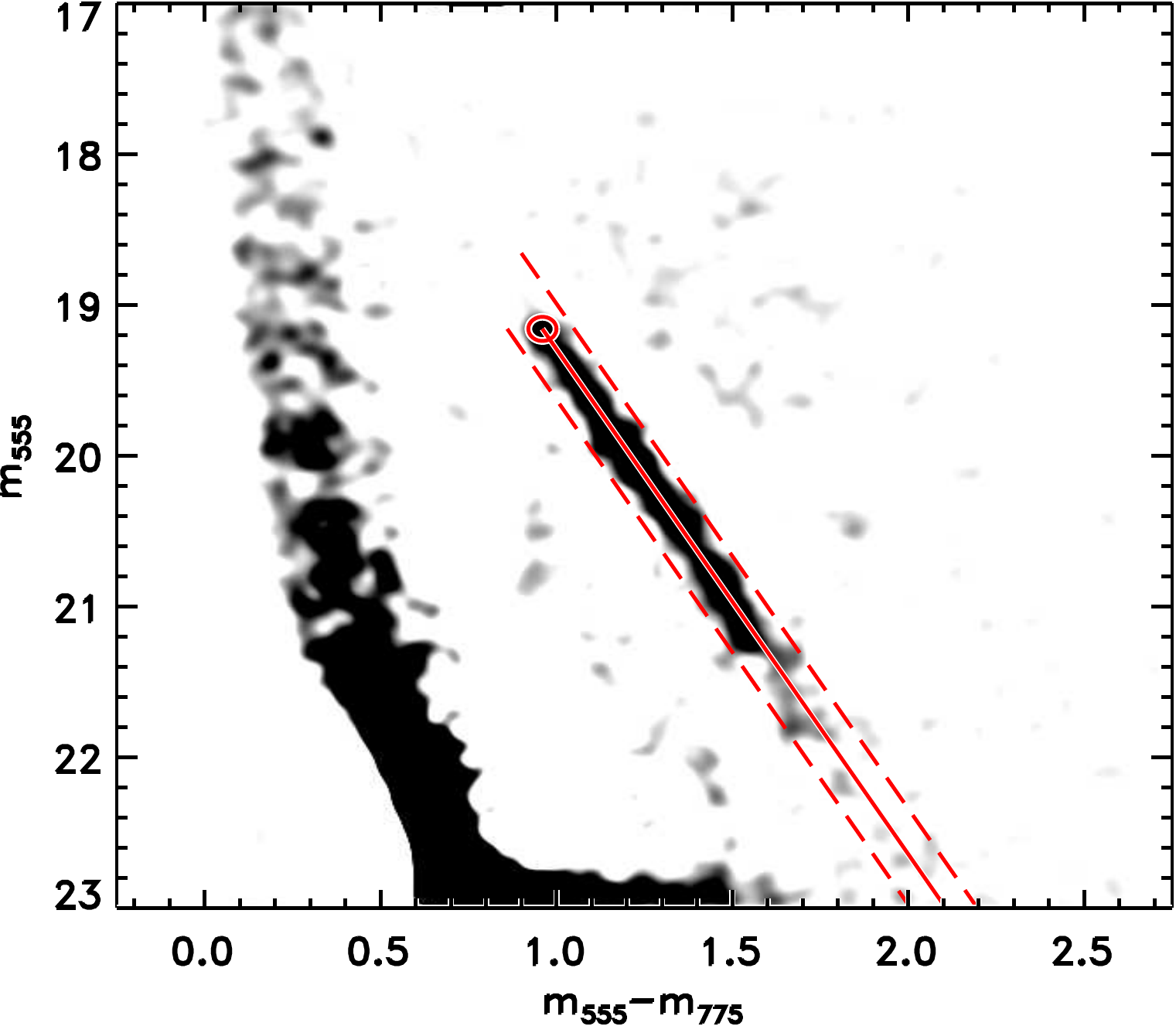}}
\caption{Same CMD as in Figure\,\ref{fig2} after unsharp masking. The
solid line traces the ridge along the elongated RC, while the ellipse
corresponds to the expected location of the un-extinguished RC. The
dashed lines define an envelope to the distribution of RC stars in the
CMD (see Section\,5).}
\label{fig4}
\end{figure}

Application of the unsharp-masking kernel to the CMD results in a an
improved definition of local density enhancements, such as the 
elongated RC sequence or the sub-structures along the UMS (see 
Figure\,\ref{fig4}). These features were already present in 
Figure\,\ref{fig2}, but they were harder to distinguish and characterise
quantitatively due to the high density of more uniformly distributed
objects around them. By reducing the contrast of the low-frequency
component in the CMD (i.e. of the points more uniformly distributed),
unsharp masking makes it easier to identify high-frequency structures
otherwise overwhelmed in the background. The location of these
overdensities in the CMD is crucial to identify the parameter space
defined by objects sharing a common evolutionary phase. Although it is
not possible to identify which individual objects are in that specific
evolutionary phase because there are potentially also other stars in 
that region of the CMD, knowing where the objects are located in the
parameter space of the CMD provides critical constraints to stellar
evolution.

An example of the local overdensities are the knots seen along the UMS,
particularly in the range $20 \la V \la 21$, which correspond to
multiple turn-on points where pre-main sequence (PMS) stars reach the MS
(Hunter et al. 1995b; Sirianni et al. 2000; Brandner et al. 2001;
Cignoni et al. 2010), suggesting the existence of separate generations
of stars in the field (see De Marchi et al. 2011a; Cignoni et al 2015).
Another example are the knots seen along the elongated RC sequence,
indicating that the density distribution of the ISM is not uniform and
that major, discrete structures are present in the field as well as
along the line of sight. We will discuss these findings in detail in a
forthcoming paper (Panagia \& De Marchi, in preparation). 

%% {  [che ci siano gnochetti lungo la linea di vista lo si vedra'  
%% solo guardando dove sono le stelle che cadono nello stesso  
%% gnocchetto e se stelle nello stesso punto del campo cadono in  
%% gnocchetti diversi, quindi scritto qui e' un atto di fede...]}.  

\begin{figure*}
\centering
%\resizebox{\hsize}{!}{\includegraphics{httpfig29.pap.pdf}
%                      \includegraphics{httpfig28.pap.pdf}
%                      \includegraphics{httpfig27.pap.pdf}}
%\resizebox{\hsize}{!}{\includegraphics{httpfig32.pap.pdf}
%                      \includegraphics{httpfig30.pap.pdf}
%                      \includegraphics{httpfig31.pap.pdf}}
  \resizebox{\hsize}{!}{\includegraphics{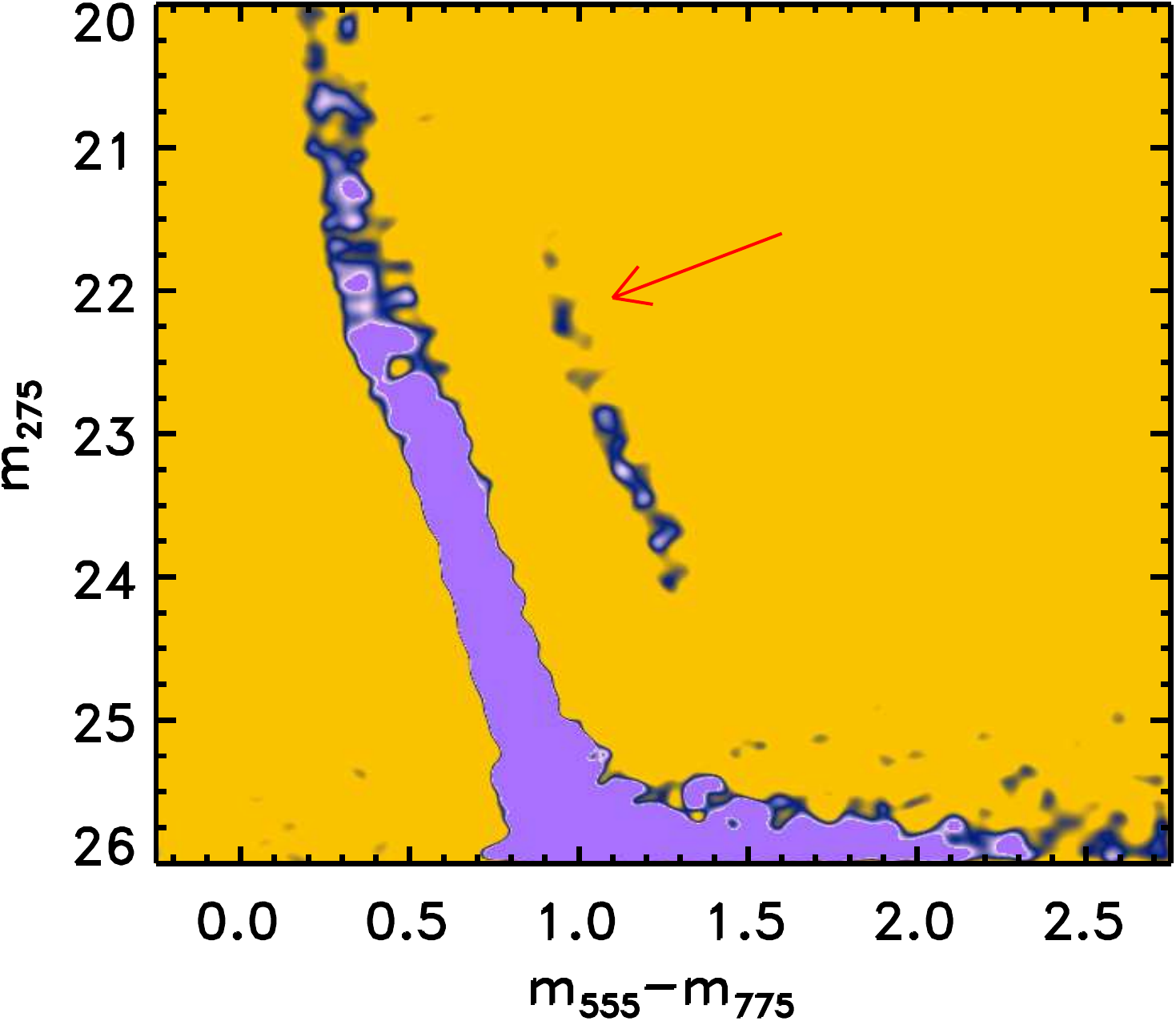}
                        \includegraphics{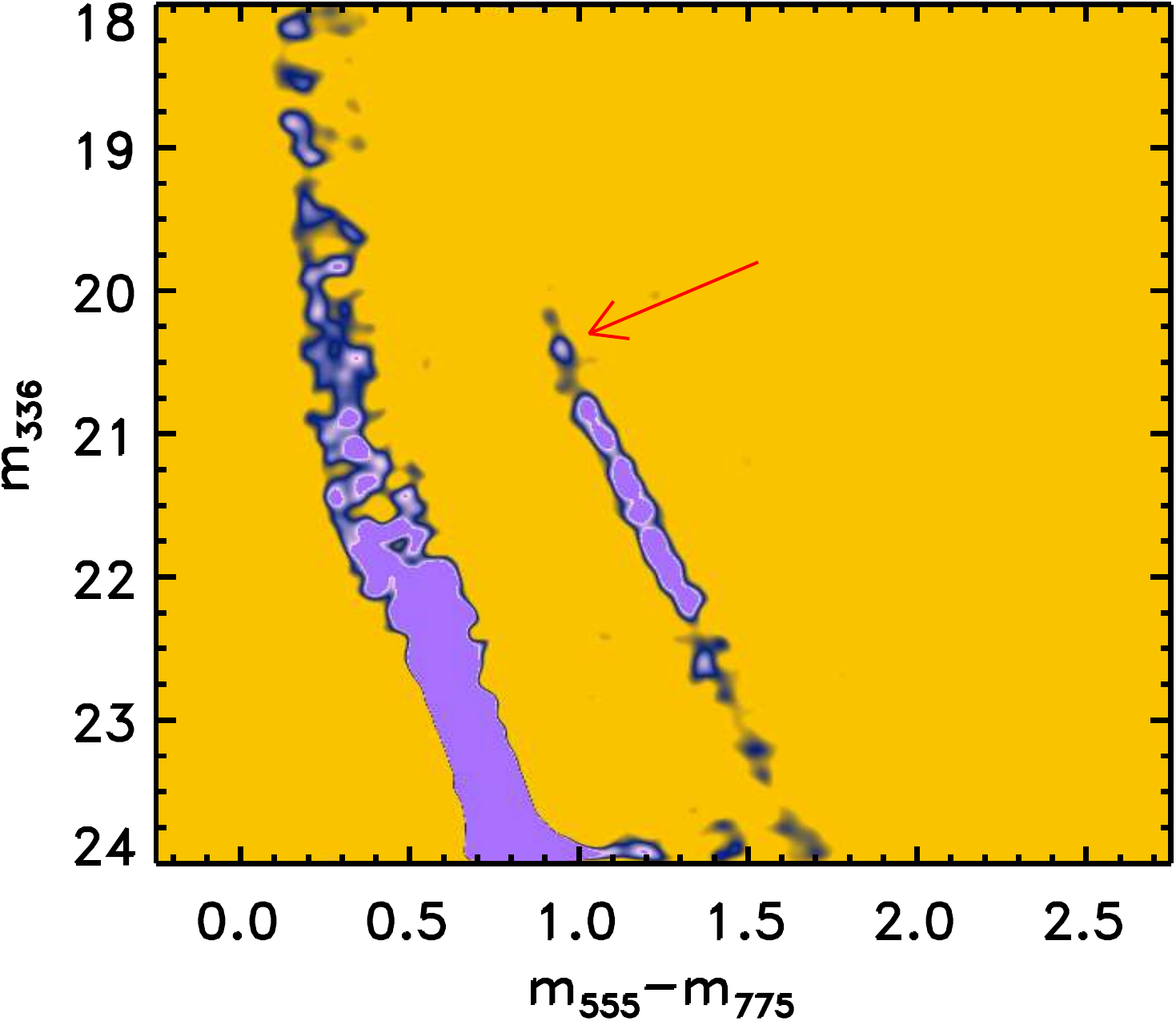}
                        \includegraphics{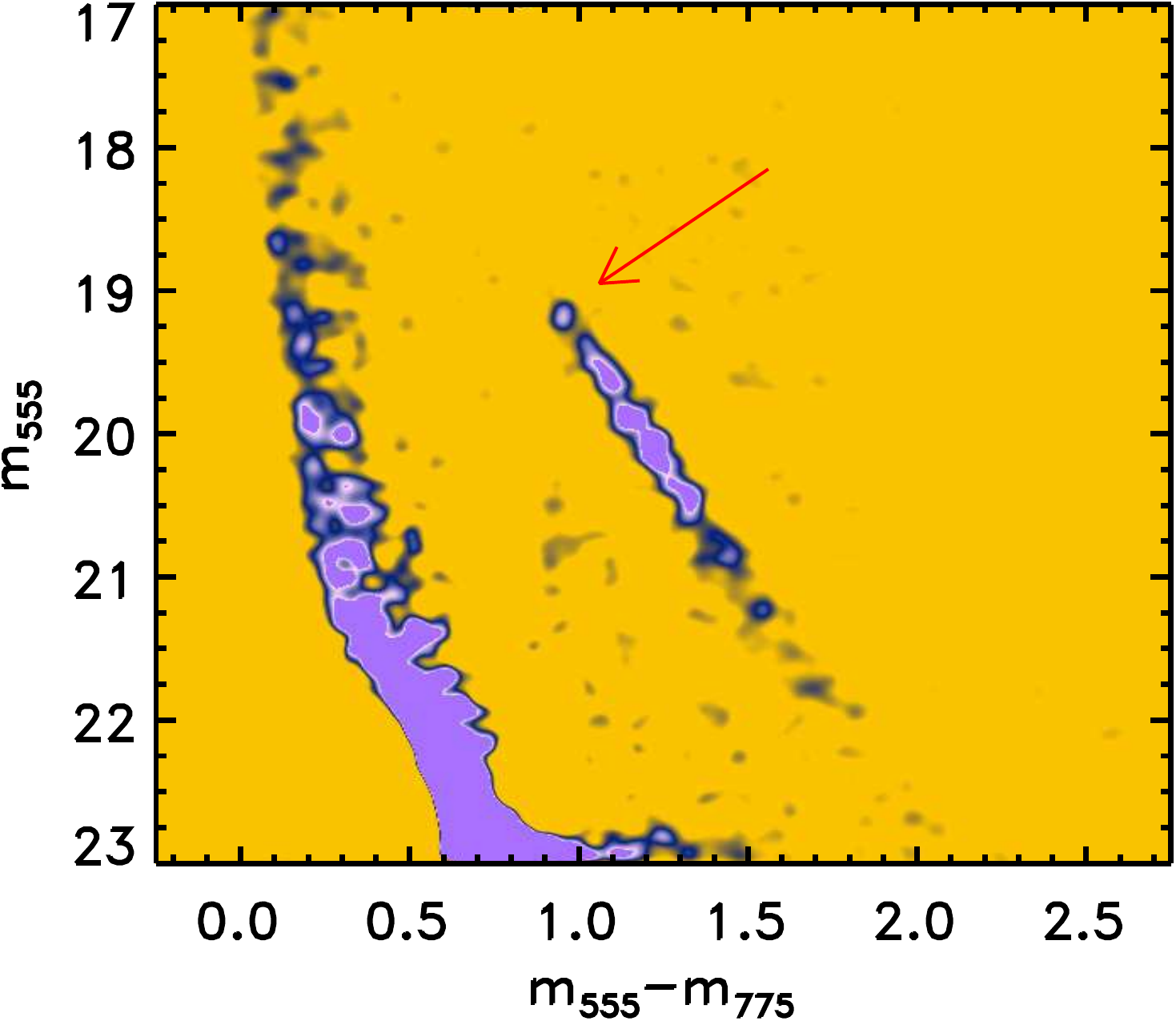}}
  \resizebox{\hsize}{!}{\includegraphics{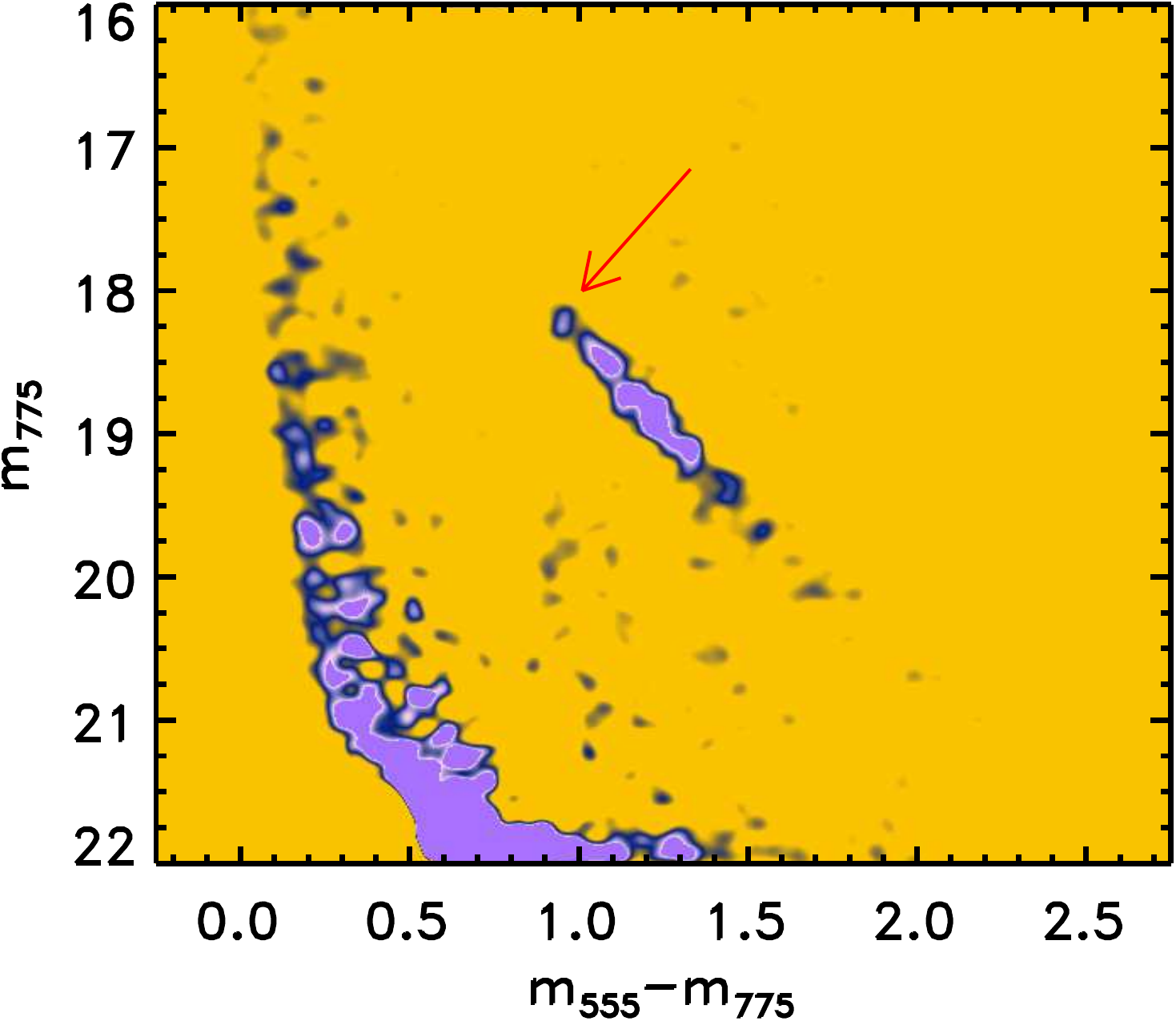}
                        \includegraphics{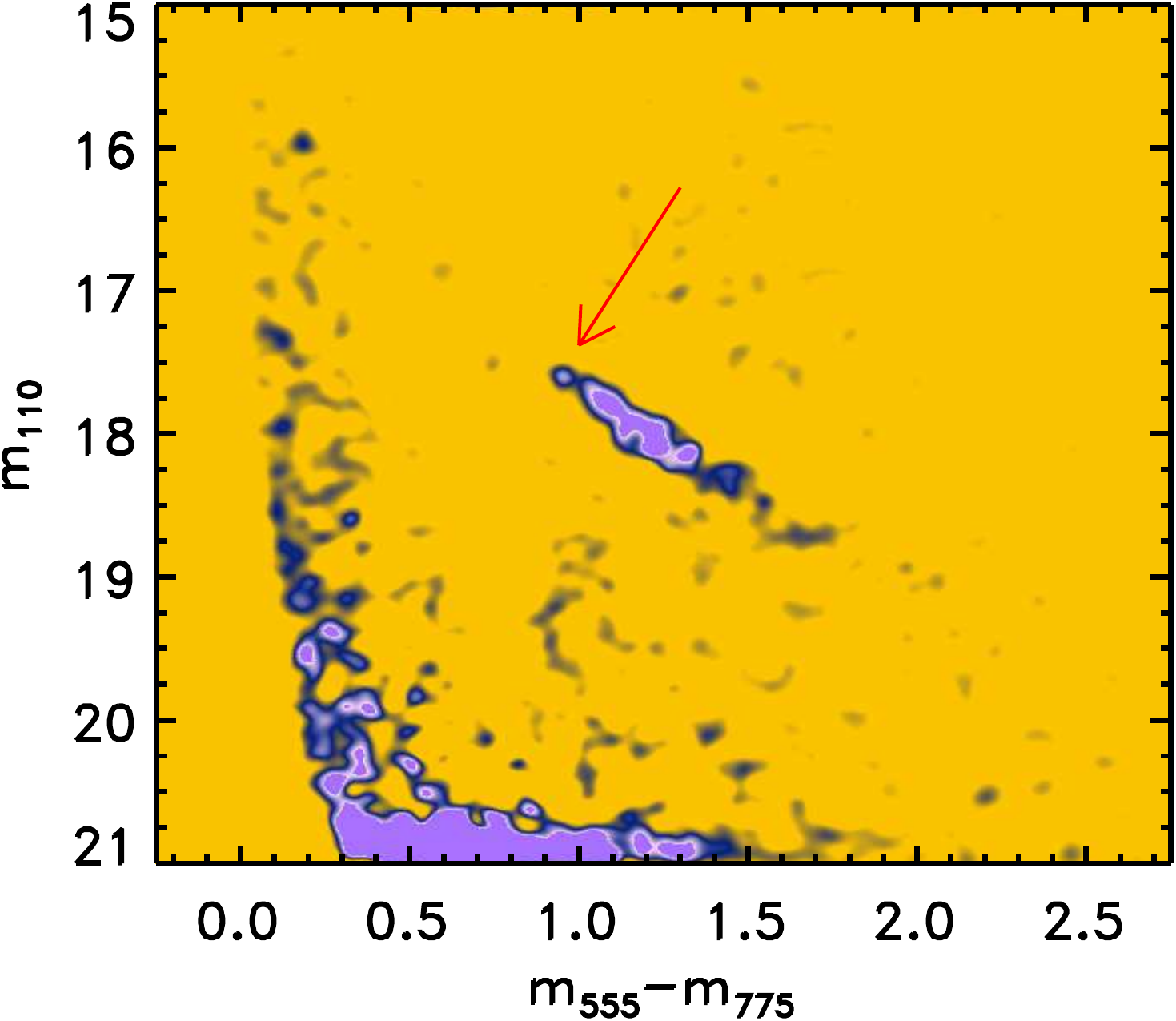}
                        \includegraphics{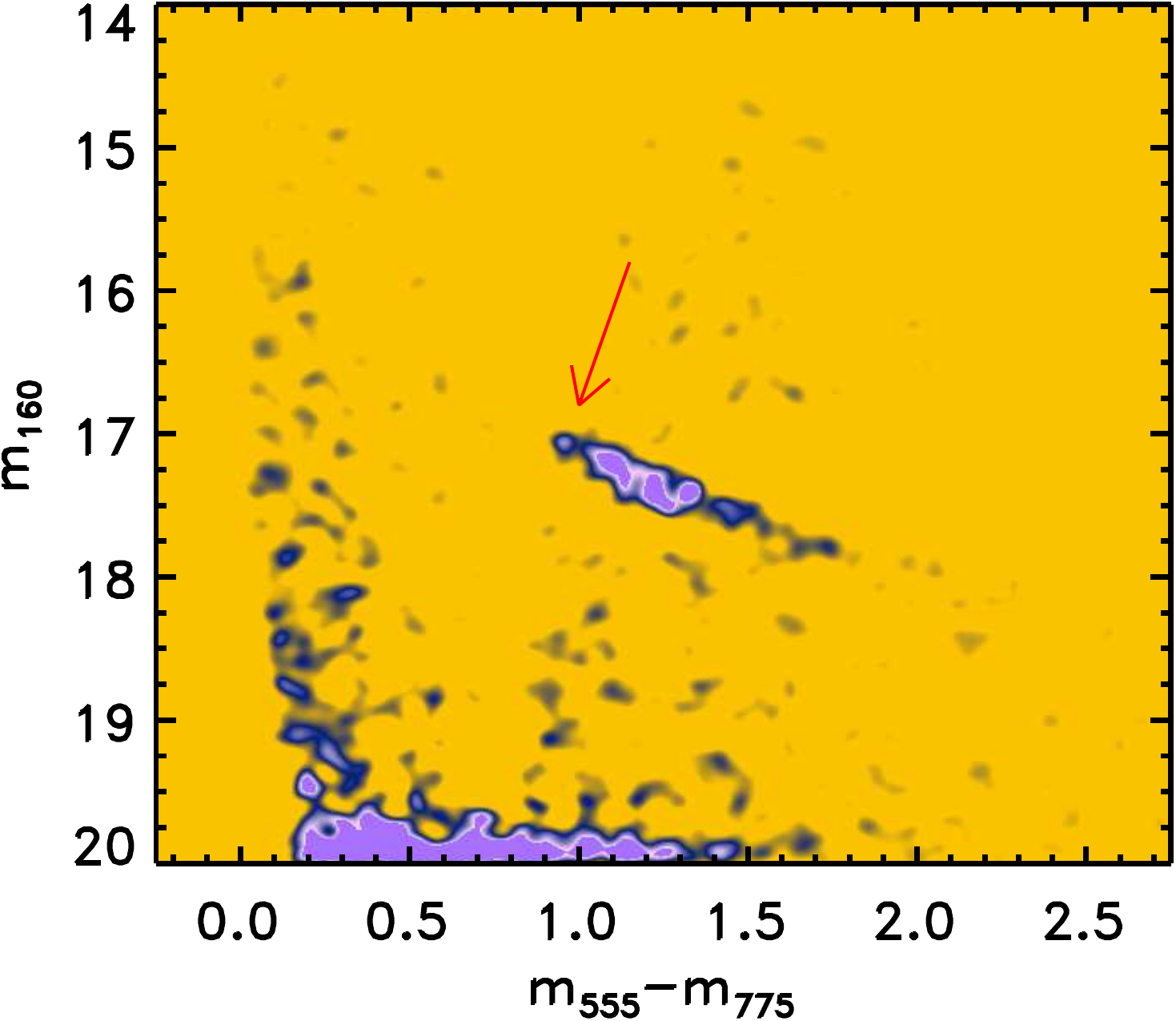}}
\caption{Unsharp masking applied to all CMDs clearly reveals the
location of the un-extinguished RC, indicated by the arrows. The slope of
the reddening vector in all bands is obtained through a linear fit to
the elongated RC.}
\label{fig5}
\end{figure*}

As regards the elongated distribution of RC stars, unsharp masking 
allows us to derive very accurately its ridge line, as shown in
Figure\,\ref{fig4}. Consequently, we can easily determine the direction
of the reddening vector. No prior knowledge of the location of the
un-extinguished RC is needed, since the head of the distribution
empirically defines the un-extinguished RC. In fact, the figure shows
that there is an excellent match between the observed head of the
elongated RC sequence and the theoretical un-extinguished RC (ellipse).
This confirms the validity of the parameter set that we had assumed,
namely the theoretical models of the RC (Girardi \& Salaris 2001;
Salaris \& Girardi 2002), the value of the foreground extinction
(Fitzpatrick \& Savage 1984) and the distance modulus (Panagia et al.
1991; Panagia 2005). These parameters are independent of one another and
independent of the measurements, so the good match seen in
Figure\,\ref{fig4} confirms the appropriateness of the adopted values. 

As mentioned above, it is important to remove from the CMD of
Figure\,\ref{fig2} possible outliers that are present in the region of
the extended RC without being RC stars. This is likely the case of the
very young PMS stars, whose broad-band colours could place them in that
CMD region. PMS objects are known to populate in large amounts the core
of 30\,Dor (e.g. De Marchi et al. 2011a), and are easily identified
through their H$\alpha$ excess emission following the method described
by De Marchi, Panagia \& Romaniello (2010) and De Marchi et al. (2011b).
Therefore, before applying unsharp masking to the CMD of
Figure\,\ref{fig2}, we have removed from the region inside the dashed
lines about 1\,\% of the stars, since they have H$\alpha$ equivalent
width $W_{\rm eq}(H\alpha) > 3$\,\AA\ and could be PMS stars (White \&
Basri 2003). The fraction is very small because, according to models of
PMS stars (e.g. Tognelli, Degl'Innocenti \& Prada Moroni 2012), only
objects younger than $\sim 0.25$\,Myr may be found in that region of the
CMD. 

% Therefore  their presence would have not affected our measurement of the
% reddening slope. {\rm GIVE RATIO AS A FUNCTION OF LOCATION, Michele's
% comment.}

% It is important to understand that unsharp masking as a tool to
% reveal ``undercover'' structures in the CMD requires certain
% conditions to be effective. Both the density (and hence the
% statistics) of the objects that one tries to reveal as well as the
% density of neighbouring objects need to be sufficiently high in
% the CMD.

\begin{table*}
\centering 
\caption{Values of the ratio $R$ between absolute ($A$) and 
selective ($E$) extinction across the field of our observations,
with the corresponding uncertainties. The effective wavelength ($\lambda$)
and wave number ($1/\lambda$) of each band are also indicated.}
\begin{tabular}{cccccccccc} 
\hline
(1) & (2) & (3) & (4) & (5) & (6) & (7) & (8) & (9) & (10) \\
\hline   
Band combination & $\lambda$ & $1/\lambda$ & $R$ & $R$ & $R$ & $R$ & $R$ & 
$R$ & $R$ \\
 & [\AA] & [$\muup$m$^{-1}$] & Whole field & North & South & Northeast & 
Northwest & Southeast & Southwest \\
\hline

$A_{275}/E(m_{555}-m_{775})$ & ~2\,712 & $3.69$ & 
 $5.15 \pm 0.38$ & $5.15 \pm 0.49$ & $5.12 \pm 0.55$ & 
 $5.07 \pm 1.01$ & $4.96 \pm 0.86$ & $4.71 \pm 0.56$ & $5.11 \pm 0.79$ \\
$A_{336}/E(m_{555}-m_{775})$ & ~3\,356 & $2.98$ &
 $4.79 \pm 0.19$ & $4.92 \pm 0.29$ & $4.69 \pm 0.23$ & 
 $5.00 \pm 0.38$ & $4.82 \pm 0.47$ & $4.55 \pm 0.40$ & $4.81 \pm 0.38$ \\
$A_{555}/E(m_{555}-m_{775})$ & ~5\,322 & $1.88$ &
 $3.35 \pm 0.15$ & $3.44 \pm 0.21$ & $3.20 \pm 0.19$ & 
 $3.47 \pm 0.14$ & $3.41 \pm 0.36$ & $3.12 \pm 0.32$ & $3.28 \pm 0.34$ \\
$A_{775}/E(m_{555}-m_{775})$ & ~7\,680 & $1.30$ &
 $2.26 \pm 0.14$ & $2.44 \pm 0.19$ & $2.12 \pm 0.17$ & 
 $2.47 \pm 0.38$ & $2.41 \pm 0.31$ & $2.03 \pm 0.30$ & $2.20 \pm 0.34$ \\
$A_{110}/E(m_{555}-m_{775})$ & 11\,608 & $0.86$ &
 $1.41 \pm 0.15$ & $1.54 \pm 0.18$ & $1.26 \pm 0.19$ & 
 $1.59 \pm 0.33$ & $1.52 \pm 0.34$ & $1.23 \pm 0.30$ & $1.30 \pm 0.43$ \\
$A_{160}/E(m_{555}-m_{775})$ & 15\,387 & $0.65$ &
 $0.95 \pm 0.18$ & $0.98 \pm 0.24$ & $0.90 \pm 0.22$ & 
 $1.00 \pm 0.49$ & $0.98 \pm 0.40$ & $0.89 \pm 0.33$ & $0.95 \pm 0.51$ \\
\hline
\end{tabular}
\vspace{0.5cm}
\label{tab2}
\end{table*}

\section{The extinction law and its properties}

\subsection{Deriving the extinction law}

{  From the CMDs shown in Figure\,\ref{fig5},} we derived the slope
of the reddening vector in all bands, as a function of the $m_{555} -
m_{775}$ colour. These slopes are in fact the ratio $R$ of absolute
($A$) and selective ($E$) extinction in the specific bands. To derive an
accurate measure of $R$ in each diagram, we used a linear fit along the
extended RC, with weights proportional to the density of objects in the
CMD after unsharp masking. The values of $R$ and corresponding
uncertainties are listed in Table\,\ref{tab2} for various areas across
the field. In addition to the full $16\arcmin\times 13\arcmin$ region,
we have measured the values of $R$ separately in the northern and
southern half fields, as well as in the four quadrants. 

As for the uncertainties, the values listed in Table\,\ref{tab2}
correspond to the dispersion around the best fit. {  Formally, the
uncertainty on the mean is a factor $\sqrt{N}$ smaller, where $N$ is the
number of stars along the extended RC between the dashed lines in
Figure\,\ref{fig4}, which amounts to $\sim 3\,500$ objects over the
whole field. Therefore, the formal uncertainties on $R$ are typically
less than $0.5\,\%$. However, this would be fully correct only if the 
filters were monochromatic and if the properties of the grains were the
same everywhere. In fact, the filters are broad bands and, as we will
show, small variations in the reddening slopes suggest slight
differences in the grain properties. {\rm Even though these differences
are small, they indicate that the extinction law is not exactly the same
across the area  and, therefore, we cannot take advantage of the
statistics because the dispersion is not only due to observational
uncertainties. Hence, we quote the  root-mean-square uncertainties
listed in Table\,\ref{tab2} since } they represent the actual extent of
the dispersion.} 

Note that the systematically larger uncertainties on the slope  $R$ in
the F275W band are due to the considerably shorter exposure times {  and
the ``red leak''} affecting this band (see Sabbi et al. 2015 and Dressel
2015 for details). 

Within the formal uncertainties listed in Table\,\ref{tab2}, the slopes
$R$ in the four quadrants are in agreement with one another. There is,
however, a small systematic variation in the SE quadrant, where $R$ is
always lower than in the rest of the field, by about $12\,\%$. The
origin of this small difference will be discussed in more detail in a
future work (De Marchi, Panagia, et al. 2015, in preparation), but it is
likely related to the smaller fraction of large  grains in this region,
compared to the rest of the field. {\rm What is particularly remarkable,
however, is the linearity of the extinction feature seen in all panels
of Figure\,\ref{fig5}. If the properties of the grains were changing with
the environment, one would expect that they should be very different in
regions of high and low extinction, but this is not what
Figure\,\ref{fig5} suggests. We will discuss this in detail in a future
paper (De Marchi, Panagia, et al. 2015,  in preparation), but it is
already clear that the linearity of the features sets stringent limits
on the spatial variation of grain  properties. }

In the following, we will derive the extinction law for the whole region
covered by the HTTP observations from the values listed in Column (4). It
is customary to express the extinction law through the ratio

\begin{equation}
R_{BV}(\lambda) \equiv \frac{A(\lambda)}{E(B-V)},
\label{eq2}
\end{equation}

\noindent  
where $A(\lambda)$ is the extinction in the specific band and $E(B-V)$
the colour excess in the canonical Johnson $B$ and $V$ bands. Since our
observations do not include a filter close in wavelength to the Johnson
$B$ band, expressing the $R$ values of Table\,\ref{tab2} as a function
of $E(B-V)$ as required by Equation\,\ref{eq2} could result in larger
uncertainties. Instead, we will express the extinction law using the
standard $E(V-I)$ colour excess, since our observations include filters
very close to these standard Johnson -- Cousin bands and interpolation
in that case is much more robust. Conversion of the measured values of
Table\,\ref{tab2} into 

\begin{equation}
R_{VI}(\lambda) \equiv \frac{A(\lambda)}{E(V-I)}
\label{eq3}
\end{equation}

\noindent
is easily done through spline interpolation (see De Marchi et al. 2014).

\begin{figure*}
\centering
%\resizebox{\hsize}{!}{\includegraphics[width=16cm]{httpfig50.pdf}}
 \resizebox{\hsize}{!}{\includegraphics[width=14cm]{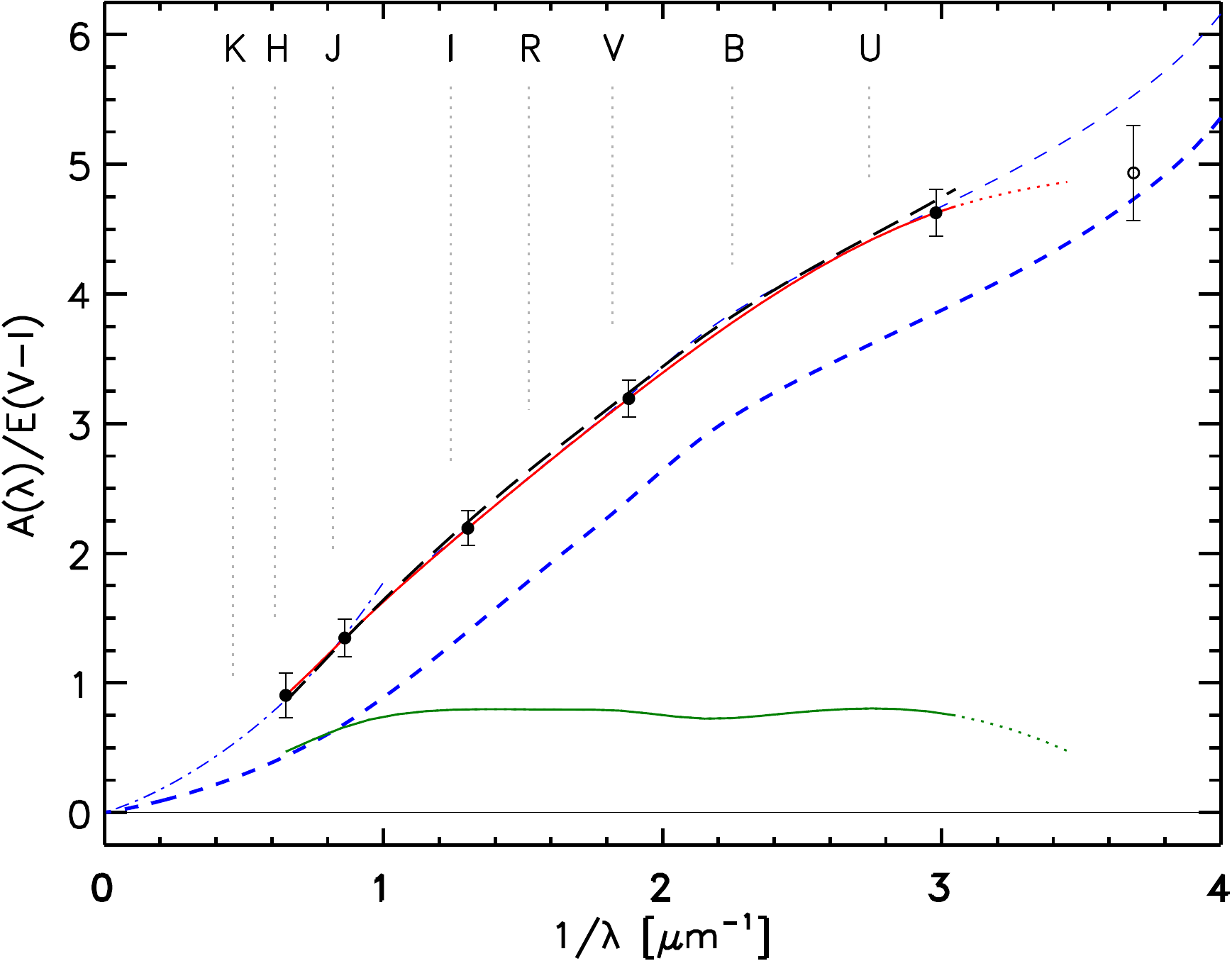}}
\caption{Extinction law. The dots show the measurements, with their
uncertainties, over the entire $\sim 16^\prime \times 13^\prime$ field,
while the red solid line is a spline interpolation through the values.
The uncertainty grows in the NUV domain due to the shallow photometry
and filter red leak in the F275W band. The long-dashed line displays the
extinction law measured by De Marchi \& Panagia (2014) within $1\farcm4$
radius of R\,136 and is in excellent agreement with that measured in
this work over the entire Tarantula nebula. The thick short-dashed line
indicates the canonical Galactic extinction law, taken from Fitzpatrick
\& Massa (1990; see also Fitzpatrick 1999) for $R_V=3.1$, corresponding
to $R_{VI}(V)=2.3$. The thin short-dashed line shows the same law,
shortwards of $\sim 1\,\muup$m shifted vertically by $0.8$ to fit the
Tarantula observations in the optical. The dot-dashed line is the
Galactic extinction law, longwards $\sim 1\,\muup$m, multiplied by a
factor of 2 to fit the measured values in the  $J$ and $H$ bands. The
green solid line is the difference between the Galactic extinction law
(thick short-dashed line) and that of the Tarantula nebula (red solid
line).}
\label{fig6}
\end{figure*}

The $R_{VI}(\lambda)$ values obtained in this way are shown in
Figure\,\ref{fig6}, for the specific wavelengths of our observations.
The dots are the values derived for the entire field and the red solid
line shows a spline interpolation through the points. As mentioned
above, the systematically shorter exposures and the red leak (Dressel
2015) in the F275W band cause a larger uncertainty on the value of
$R_{VI}(\lambda)$ in the near ultraviolet (NUV). For this reason, we
have indicated with a dotted line the spline interpolation in that
wavelength range. 

For easier comparison with previous works, the interpolated
$R_{VI}(\lambda)$ values in the classical Johnson--Cousin bands, at the
wavelengths marked by the vertical dotted lines in the figure, are also
listed in Table\,\ref{tab3}. Note that the value for the $K$ band is
actually an extrapolation and, therefore, is indicated in italics in the
table. Besides the value of  $R_{VI}(\lambda)$, in Table\,\ref{tab3} we 
provide for reference also the values of $R_{BV}(\lambda)$ as per
Equation\,\ref{eq2}, but we stress again that they are less accurate
because of the lack of observations near the wavelengths of the $B$
band, as mentioned above.

\begin{table}
\centering 
\caption{Interpolated values of $R_\lambda$ for the most common bands,
as a function of both $E(V-I)$ and $E(B-V)$. The latter is less accurate
because of the lack of $B$-band observations in our photometry.
The table also gives the effective wavelengths ($\lambda$) and wave
numbers ($1/\lambda$) of the filters, the value of $R_\lambda^{MW}$ for
the canonical extinction law in the diffuse Galactic ISM, and the
difference between the latter and our measurements. All values are given
for the specific monochromatic effective wavelength as indicated,
ignoring the width of the filters. The values for the $K$ band are
extrapolated.}
\begin{tabular}{cccccc} 
\hline
Band & $\lambda$  & $1/\lambda$  &  $R_{VI}(\lambda)$ &
$R_{VI}^{MW}(\lambda)$ & $R_{VI}^{diff}(\lambda)$\\
 & [\AA] & [$\muup$m$^{-1}$] & & & \\
\hline
$U$ & ~3\,650 &  $2.74$ & $4.41\pm0.18$ & $3.61$ & $0.80$ \\
$B$ & ~4\,450 &  $2.25$ & $3.78\pm0.15$ & $3.05$ & $0.73$ \\
$V$ & ~5\,510 &  $1.82$ & $3.09\pm0.15$ & $2.30$ & $0.79$ \\
$R$ & ~6\,580 &  $1.52$ & $2.58\pm0.13$ & $1.78$ & $0.80$ \\
$I$ & ~8\,060 &  $1.24$ & $2.09\pm0.17$ & $1.29$ & $0.79$ \\
$J$ & 12\,200 &  $0.82$ & $1.26\pm0.18$ & $0.63$ & $0.63$ \\
$H$ & 16\,300 &  $0.61$ & $0.84\pm0.12$ & $0.40$ & $0.44$ \\
$K$ & 21\,900 &  $0.46$ & $\mathit{0.52\pm0.08}$ & $0.26$ & 
  $\mathit{0.26}$ \\
\hline
Band & $\lambda$  & $1/\lambda$  &  $R_{BV}(\lambda)$ &
$R_{BV}^{MW}(\lambda)$ & $R_{BV}^{diff}(\lambda)$\\
 & [\AA] & [$\muup$m$^{-1}$] & & & \\
\hline
$U$ & ~3\,650 &  $2.74$ & $6.39\pm0.28$ & $4.75$ & $1.64$ \\
$B$ & ~4\,450 &  $2.25$ & $5.48\pm0.24$ & $4.04$ & $1.44$ \\
$V$ & ~5\,510 &  $1.82$ & $4.48\pm0.24$ & $3.04$ & $1.44$ \\
$R$ & ~6\,580 &  $1.52$ & $3.74\pm0.20$ & $2.35$ & $1.39$ \\
$I$ & ~8\,060 &  $1.24$ & $3.03\pm0.26$ & $1.70$ & $1.33$ \\
$J$ & 12\,200 &  $0.82$ & $1.83\pm0.28$ & $0.83$ & $1.00$ \\
$H$ & 16\,300 &  $0.61$ & $1.22\pm0.18$ & $0.53$ & $0.69$ \\
$K$ & 21\,900 &  $0.46$ & $\mathit{0.75\pm0.11}$ & $0.34$ & 
  $\mathit{0.41}$ \\
\hline      
\end{tabular}
\vspace{0.5cm}
\label{tab3}
\end{table}

\subsection{Comparison with previous works}

In Figure\,\ref{fig6} we also show as a long-dashed line the
extinction curve measured by De Marchi \& Panagia (2014) with the same
method in a smaller region of the Tarantula nebula, namely the central
$1\farcm4$ radius around R\,136. The agreement between the extinction
curves over the whole Tarantula field and the R\,136 area is very good,
over the common wavelength range. This is remarkable because De Marchi
\& Panagia (2014) covered a much smaller region (sampling only about 140
RC stars instead of the roughly 3\,500 objects in this work), and the
two studies did not use the exact same set of filters: instead of F775W,
De Marchi \& Panagia (2014) used F814W. The excellent agreement
indicates that the method is solid.

The thick short-dashed line in Figure\,\ref{fig6} displays the
canonical extinction law for the Galactic diffuse ISM, from the work of
Fitzpatrick \& Massa (1990; see also Fitzpatrick 1999) for
$R_V=R_{BV}(V)=3.1$. When expressed in units of  $E(V-I)$, as per
Eq.\,\ref{eq3}, this corresponds to $R_{VI}(V)=2.3$, which is a value
significantly smaller than the $R_{VI}(V)=3.09 \pm 0.15$ that we measure
in the Tarantula (see Table\,\ref{tab3}). 

Gordon et al. (2003) have studied the extinction properties towards
eight different lines of sight associated with the LMC\,2 Superbubble
near the Tarantula nebula. Unfortunately, none of them are included
within the field of view of our observations, so it is not possible to
make a direct comparison. The closest objects are Sk\,$-68^\circ 140$,
located some $9^\prime$ NE of R\,136, and Sk\,$-69^\circ 228$, about
$17^\prime$ SW of it. These and the other more distant lines of sight
probed by Gordon et al. (2003) sample regions of considerably more
diffuse ISM than those characteristic of the Tarantula nebula itself. It
is, therefore, not surprising that the $R_V$ values measured
spectroscopically by Gordon et al. (2003) for these stars, e.g.
$R_V=R_{BV}(V)=3.27 \pm 0.24$ for Sk\,$-68^\circ 140$ or
$R_V=R_{BV}(V)=3.35 \pm 0.33$ for Sk\,$-68^\circ 228$, do not match the
value $R_{BV}(V)=4.48 \pm 0.17$ measured by De Marchi \& Panagia (2014)
for the R\,136 region and confirmed by these observations for the
Tarantula nebula at large. In fact, at optical wavelengths the
extinction curve obtained by Gordon et al. (2003) from the LMC\,2
Supershell sample is very similar to the Galactic extinction law of
Fitzpatrick \& Massa (1990; see De Marchi \& Panagia 2014 for a direct
comparison), which we already concluded does not agree with the
extinction curve in the Tarantula (see Figure\,\ref{fig6}).

Combining HST optical photometry (De Marchi et al. 2011a) with 
spectroscopy and near-infrared (NIR) photometry from the VLT-FLAMES
Tarantula Survey (Evans et al. 2011), Ma\'{\i}z Apell\'aniz et al.
(2014) derived the extinction law for the R\,136 cluster. Their study of
a sample of 83 stars of spectral types O and B with the Bayesian code
CHORIZOS (Ma\'{\i}z Apell\'aniz 2004) concluded that, inside the
cluster, the value of $R_{BV}(V)$ is larger than in the Galactic ISM.
They find $R_{BV}(V)=4.4 \pm 0.7$ when all 83 objects are considered.
Limiting the sample to the 50 objects with the smallest uncertainties,
the same average value is found but the spread is reduced, namely,
$R_{BV}(V)=4.4 \pm 0.4$. These values are in excellent agreement with
those shown in Table\,\ref{tab3} for the entire Tarantula nebula.  

An immediate implication of our extinction law for the young stars in
the Tarantula nebula is that their intrinsic brightness has been so
far systematically underestimated. With a median colour excess for UMS
stars of $E(V-I)\simeq 0.5$, the difference between the Tarantula and
Galactic extinction laws implies that one would obtain systematically
fainter intrinsic fluxes by a factor of $\sim 1.5$, on average, and by
more than a factor of 2 for the most extinguished 10\,\% of the stars.
As an example, Crowther et al. (2010) derived new luminosities and
masses for the most massive members of R\,136, correcting their
photometry with the extinction law of Fitzpatrick \& Savage (1984),
namely $R_{BV}^{\rm 30Dor}=3.7$. Even approximating this value to 4, as
Crowther et al. (2010) have done, the difference with our slope is large
($R_{BV}^{\rm 30Dor}=4.48 \pm 0.24$, see Table\,\ref{tab3}), and implies
that the luminosities and masses of the stars are in fact considerably 
higher than Crowther et al'.s (2010) estimates. For instance, the
luminosity of R\,136c grows from $\log L=6.75$ to $\log L=6.9$, which
according to their models brings the mass of the star from 220\,\Msolar\
to more than 300\,\Msolar. 

% Crowther et al. (2010, MNRAS, 408, 731) dicono che prendono la ricetta
% per l'estinzione da Fitzpatrick & Savage, cioe' per le loro 4 stelle
% hanno: Av = 0.07 * Rv(MW) + 0.16 Rv(LMC) + (0.25 - 0.45) * Rv(30 Dor)
% Usando il valore tipico 0.35 per E(B-V) per questi oggetti, trovo
% Av=2.14 visto che prendono Rv(MW)=Rv(LMC)=3.2 e Rv(30Dor)=4. Se per
% Rv(30Dor) prendiamo il nostro valore, 4.5, troviamo Av=2.45. La
% differenza e' un fattore 1.33 in luminosita'. Non grandissimo ma nemmeno
% trascurabile. Certo, se in realta' il E(B-V) della stella e' piu' vicino
% a 0.45 (questo e' il caso di R136c) le differenze aumentano ancora a un
% fattore 1.45. 
% R136c non e' il piu' massiccio degli oggetti che studiano, ma gli
% assegnano una massa di ~220 Msolar. Con la giusta legge di estinzione,
% la luminosita' diventerebbe come quella della stella piu' brillante,
% R136a1, e la massa secondo i loro modelli sarebbe pure di oltre 300
% Msolar, come loro stimano per R136a1. 

\subsection{Grain properties}

As De Marchi \& Panagia (2014) concluded, the extinction law in these
regions is {  ``flatter''} than in the Galactic ISM, i.e. less steep
in logarithmic terms. As one can see from Figure\,\ref{fig6}, at optical
wavelengths {  in linear terms} the extinction law in these regions is
almost exactly parallel to the Galactic curve. The thin short-dashed
line is the portion of the standard Galactic law shortwards of
$1\,\muup$m with a vertical offset of $0.8$ and it matches the measured
extinction curve surprisingly well (the match is in fact so good that
the thin short-dashed line is often hard to discern). 

The difference between the Galactic extinction law and the one in the
Tarantula is shown by the green solid line in Figure\,\ref{fig6}. At
wavelengths shorter than $\sim 1\,\muup$m the difference is practically
constant (see also Table\,\ref{tab3}). A noticeable feature is the small
dip at $1/\lambda \simeq 2.2\,\muup$m$^{-1}$ or $\lambda \simeq
4\,550\,\muup$m, which is a likely consequence of the lack of $B$-band
observations in our photometry. While the Galactic extinction law
features a small knee at this wavelength (see thick short-dashed line),
our interpolation is rather smooth in this range since we have no data
points between the F336W and F555W filters. Also, we do not regard as
significant the apparent decline of the curve in the NUV because of the
larger photometric uncertainties and filter red leak at these
wavelengths, as mentioned above.

The practically constant difference between the Galactic and Tarantula
extinction curves in the optical indicates that the dust is in fact of
the same or similar type but that in the Tarantula nebula there is an
additional component. Since in the optical the contribution of this
component is ''grey'', i.e. it does not appear to depend on the
wavelength, its most likely origin is the presence of a larger fraction of
large grains than in the diffuse ISM in the Galaxy and LMC. This is the
accepted interpretation for the high ratios of total-to-selective
extinction observed in some Galactic environments (see e.g. Strom, Strom
\& Yost 1971; Jones 1972; Dunkin \& Crawford 1998; Skorzynski, Strobel
\& Galazutdinov 2003).

The NIR domain provides further indication that, except for the extra 
grey component, the extinction law in the Tarantula nebula is similar to
that of the diffuse ISM in the Galaxy or LMC. At wavelengths longer than
$\sim 1\,\muup$m, { the Tarantula extinction law tapers off as $\sim
\lambda^{-1.7}$, following almost exactly the observed properties of the
Galactic extinction law (e.g. Cardelli, Clayton \& Mathis 1989; Wang et
al. 2013).} The dot-dashed line shown in Figure\,\ref{fig6} is the
portion of the Galactic extinction law longwards of $1\,\muup$m
multiplied by a factor of 2, and it offers a remarkably good fit to our
observations in the $J$ and $H$ bands. Therefore, there is no reason to
believe that the nature of the Tarantula grains {\rm should} be
drastically different from that of the diffuse Galactic ISM.

\subsection{Role of large grains}

A detailed analysis of the grain properties as a function of the
location inside the nebula will be presented in a forthcoming work (De
Marchi, Panagia, et al. 2015, in prep.). However, as De Marchi \&
Panagia (2014) have already pointed out, simple considerations can
provide valuable insights into the properties of the additional dust
component present in the 30\,Dor regions. 

It is well known ({\em e.g.,} van de Hulst 1957; Greenberg 1968; Draine 
\& Lee 1984) that, at wavelengths short enough, the extinction {
(= absorption + scattering) cross section of a grain of radius $a$ tends
asymptotically to the geometric cross section { $\sigma_{\rm geom} =
2 \, \pi \, a^2$}. At longer wavelengths the cross section is smaller 
than $\sigma_{\rm geom}$ and becomes proportional to the grain volume.}
Conveniently enough, the transition occurs approximately at $\lambda_0
\sim 2\,\pi\,a$ and, for a given grain size, one would expect a sort of
a step function behaviour with the transition occurring rapidly around
$\lambda_0$. To account for the observed MW extinction law's steady
increase with wave number over a wide wavelength range ($0.2\,\muup$m$
\la \lambda  \la 5\,\muup$m), it is generally assumed that there is a
distribution of grain sizes of the type $f(a) \propto a^{-\beta}$, with
$\beta \simeq 3.5$ and the grain radius $a$ ranging from $a_{\rm min}
\sim 0.01\,\muup$m to $a_{\rm max} \sim 0.2\,\muup$m (Mathis, Rumpl \&
Nordsieck 1977; Draine \& Lee 1984). 

With $\beta \simeq 3.5$, at wavelengths longer than $2\,\pi\,a_{\rm max}$
the extinction is dominated by the largest grains and is proportional to
the total mass in grains. Taking the Galactic extinction law as a
reference template, the fact that in the NIR the absolute value of the
extinction in the Tarantula is about twice as large as it is in the MW
(see Figure\,\ref{fig6}) implies that the mass fraction in large grains
is about twice as high as in the MW. Therefore, the extinction law
inside the Tarantula nebula can be represented with the sum of two
components: one being the standard Galactic extinction law and the other
being made up only of large grains, which are similar in type to those
found in the diffuse Galactic ISM. 

De Marchi \& Panagia (2014) concluded that, for the central regions of
30\,Dor, the most likely origin for the higher relative abundance of
large grains is the selective injection of ``fresh'' large grains into
the MW mix. The same conclusions can now be extended to the Tarantula
nebula at large. The two other ways to explain an excess of large grains
would be selective destruction of small grains, or selective
condensation of material on the surface of small grains, but both would
imply a decrease in the number of small grains and hence an extinction
law that is flatter than the MW's at UV wavelengths. In fact,
measurements towards the stars of the Magellanic Clouds reveal a steeper
rise in the UV extinction curve compared to MW objects ({\em e.g.}
Fitzpatrick 1998 and references therein). Note that, as mentioned above,
the apparent decline of the curve at NUV wavelength in
Figure\,\ref{fig6} is not significant, due to the large photometric
uncertainties and filter red leak at those wavelengths.

The selective addition of new large grains to the mix can easily account
for the presence of the extra grey component in the extinction curve,
without conflicting against measurements at other wavelengths. {
Actually, Gall et al. (2014) recently revealed rapid formation of large,
$\muup$m-size dust grains in the dense circumstellar medium around
SN\,2010jl in the metal-poor galaxy UGC\,5189 (Newton \& Puckett 2010).
Their observations with the {\em Very Large Telescope} reveal that the
extinction curve around the supernova evolves rapidly and turns into a
mix of grey-extinction dust grains and MW dust grains. The extinction
contribution of the grey dust is about 40\,\% in the $V$ band. Also in
the LMC, } recent {\em Herschel} and {\em ALMA} observations of
SN\,1987A (Matsuura et al. 2011; Indebetouw et al. 2014) indicate that a
substantial amount ($> 0.4$\,\Msolar) of large grains ($> 0.1\,\muup$m)
is being produced in the ejecta. { A similar amount of dust is
expected in SN\,2010ji if the dust production continues to follow the
trend observed so far (Gall et al. 2014). These recent findings make
}injection of large grains by supernova explosions an exciting
possibility for the extinction law in the Tarantula as well. {
Indeed, very large, grey dust grains recently received much attention in
the literature as there is ample evidence for such $\muup$m-sized grains
in the Galactic ISM ({\em e.g.}, Wang, Li \& Jiang 2015a)}.

Star formation has been active for at least 30\,Myr in the Tarantula
nebula, and possibly longer, as witnessed by the presence of both young
and older generations of stars in NGC\,2070 (Walborn \& Blades 1997; De
Marchi et al. 2011a; Cignoni et al. 2015), in Hodge\,301 (Grebel \& Chu
2000), and in NGC 2060 (Mignani et al. 2005). If all type\,II supernova
explosions result in an output comparable to that of SN\,1987A { and
SN\,2010jl,} the excess of large grains should have built up considerably
over time and will reach a peak after about $\sim 50$\,Myr, which is the
lifetime of the 8\,\Msolar\, stars at the lower mass limit of supernova
type II progenitors. Even though the large grains are eventually
destroyed in the hot gas behind shock fronts in supernova remnants
(Draine 2009; Dwek \& Scalo 1980; Dwek 1998), with a total mass in
excess of $10^5$\,\Msolar\, in this starburst region (Bosch et al. 2001;
Andersen et al. 2009) the expected supernova rate is above
$10^{-4}$\,yr$^{-1}$ (Cervi\~no et al. 2001), implying a sustained
injection of large grains into the ISM. As an order of magnitude, one
would expect in a typical 10\,Myr time frame about 1\,000 type II SNe,
corresponding to up to $\sim 400$\,\Msolar of large grains. With the
quoted total mass ($\sim 10^5$\,\Msolar) and metallicity ($Z\simeq
0.007$; e.g. Hill, Andrievsky \& Spite 1995; Geha et al. 1998) for these
regions, the resulting mass in large grains compares favourably with the
expected $\sim 50\,\%$ fraction of metals locked in grains ({\em e.g.}
Savage \& Sembach 1996).

To confirm whether this interpretation of the observed extinction law is
indeed correct and to understand whether differences with the diffuse
Galactic ISM are mainly in the fraction of large grains, further studies
are required. Spectroscopic UV observations of early-type stars in
suitable locations inside the Tarantula nebula in the range $\sim
1\,200 -  3\,500$\,\AA are needed to probe the distribution of small
grains and to measure in which proportion they are present. These
observations are possible with the {\em Cosmic Origin Spectrograph}
(COS) on board the HST.

\subsection{Evolution of the extinction properties}

So far we have seen that, across the entire Tarantula nebula, the
extinction law {  implies a steeper reddening vector} than in the
diffuse Galactic ISM: with $R_V \simeq 4.5$, the reddening vector in the
CMDs is consistently $\sim 50\,\%$ steeper. Since this is not the case
in more diffuse ISM regions in the LMC, the extinction must be related
to the intense star formation witnessed by the Tarantula complex.  At
the same time, the properties of the extinction law, and hence those of
the grains, show relatively small variations across the $\sim 16^\prime
\times  13^\prime$ region that we studied. Indeed, a value of $R_V\simeq
4.5$ is also found in regions devoid of massive stars or in hot X-ray
super-bubbles (Wang \& Helfand 1991), where $A_V$ is generally lower
(see Sabbi et al. 2015). Therefore, this effect is not limited just to
the regions of most recent star formation, but also to those where star
formation peaked some 20--30\,Myr ago (e.g., Hodge\,301 and NGC\,2060).

If the overabundance of large grains is due to injection by type II
supernovae, as in the case of  SN\,1987A (Matsuura et al. 2011;
Indebetouw et al. 2014) { and SN\,2010jl (Gall et al. 2014)}, the ISM
enrichment in large grains will be progressive. It will begin with the
explosion of the most massive progenitors of the first generation of
stars, a few Myr into the  star-formation episode, and will continue to
increase for about $\sim 50$\,Myr, i.e. the lifetime of the 8\,Myr stars
at the lower mass limit of type II progenitors. At this stage, the
excess of large grains should be highest and it will begin to decrease
progressively as the grains are destroyed in the environment. They must
be relatively easy to destroy, since the regions around the Tarantula
show a rather standard extinction curve (Gordon et al. 2003). This would
be easy to understand if the new large grains were mostly made of ices,
which sublimate at low temperatures without affecting appreciably the
underlying grain distribution. { A grey extinction component caused
by $\muup$m-size ice grains is also compatible with the mid-IR
extinction properties of the Galactic ISM (Wang, Li \& Jiang 2015b). }
Constraints on the timescale of these phenomena can be set by comparing
the extinction properties and ages of the populations inside the
Tarantula nebula with those in the surrounding regions.

{\rm Obviously, our findings can have important implications for the
study of the star formation properties in galaxies. Beyond the nearest
Universe, star formation properties such as the star formation rate or
stellar masses are derived from diagnostics of HII regions (e.g.
Kennicutt 1998). Their integrated colours and spectra are dominated by
the energy of  massive stars and are significantly affected by
extinction, i.e. by both the amount and the properties of dust grains.
As we have concluded, in regions undergoing massive star formation the
properties of the dust grains appear to change from those characteristic
of the diffuse Galactic ISM. Even though the changes might be short
lived and might last only some $\sim 50$\,Myr, they affect the HII
regions when these are most easily detectable in distant
galaxies. Therefore, assuming typical ISM conditions in these regions
could result in severely inaccurate total masses and star formation
rates. If the $R_V=4.5$ value measured in the Tarantula nebula is anywhere
typical of massive star forming regions and the reddening is high,
assuming the classical $R_V\simeq 3$ value could result in fluxes that
are about a factor of 2 too faint. The outcome would be a seriously
underestimated star formation rate leading to a distorted view of the
formation and of the chemical evolution of galaxies (e.g. Matteucci
2012). }

\section{Extinction across the Tarantula nebula}

In this section we study the reddening distribution across the field in
order to derive a reddening map, and discuss how to use it to correct
the photometry of individual stars. Having determined the slope of the
reddening vector in all observed bands (Table\,\ref{tab2}), we can
measure the total extinction towards objects whose nominal location in
the CMD can be determined unambiguously, using them as probes of the
extinction along their respective lines of sight. These include not only
RC stars, but also the objects in the UMS, since that portion of the CMD
is not shared with stars in other evolutionary phases. 

It is crucial to understand that the extinction map resulting from this
collection of lines of sight is by definition a two-dimensional
projected distribution. On the other hand, in order to correct the
photometry of other objects in the field one would need a
three-dimensional distribution, both of the stars and of the absorbing
material, to account for their location along the line of sight. One
could be tempted to interpolate between lines of sight, but this would
not necessarily result in a more accurate photometry. In fact, it is
likely to introduce larger uncertainties. As we will show, one has
instead to use additional information (e.g. spatial distribution) to
determine a meaningful correction for stars other than the reddening
probes. 

\begin{figure}
\centering
%\resizebox{\hsize}{!}{\includegraphics[width=16cm]{httpfig34all.pdf}}
 \resizebox{\hsize}{!}{\includegraphics[width=16cm]{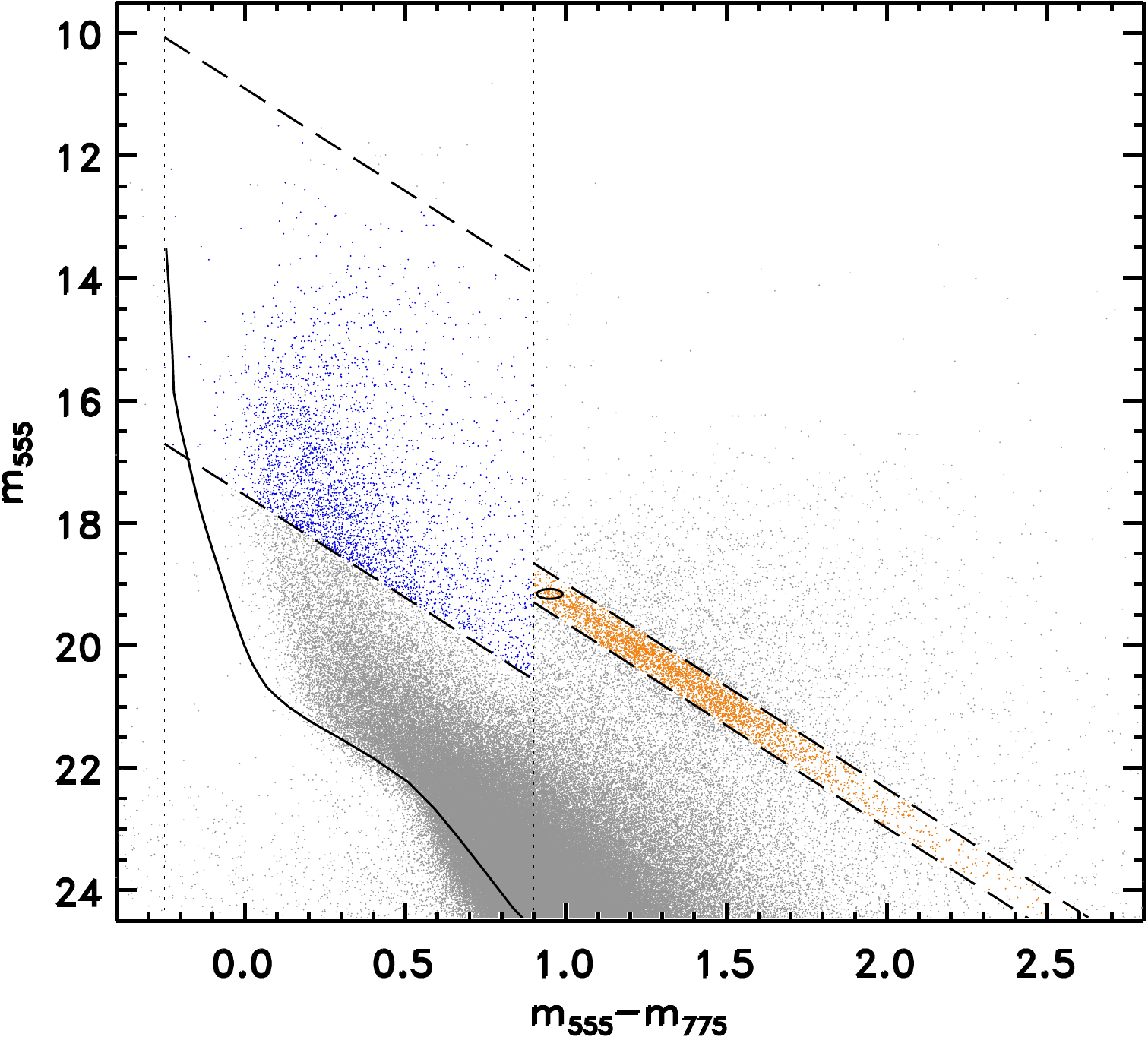}}
\caption{The stars used as reddening probes are those between the dashed
lines, which are parallel to the direction of the reddening vector in
these bands.}
\label{fig7}
\end{figure}

\subsection{Reddening distribution}

The objects serving as reddening probes are those between the dashed
lines shown in the CMD of Figure\,\ref{fig7}. For the RC, we have used
as limits the envelopes around the extended RC after unsharp masking
(see Figure\,\ref{fig4}), considering only objects redder than 
$m_{555}-m_{775}=0.9$ since that is the minimum colour consistent with
the uncertainty around the nominal RC colour in these bands, taking into
account the Galactic extinction component towards the LMC (see
Table\,\ref{tab1}). We have further restricted the selection to stars
{  more massive than 10\,\Msolar} with combined photometric uncertainty
$\delta_2 \le 0.1$, and have excluded objects with $W_{\rm eq}(H\alpha) >
3$\,\AA\ since they could be PMS stars (see Section\,3). In total, about
3\,500 objects were selected in this way, corresponding to a projected
average RC stars density of $\sim 20$\,arcmin$^{-2}$. To guarantee a
similar density of UMS stars over the same field, we selected a somewhat
larger number of objects with colours $-0.25 < m_{555}-m_{775} <0.9$,
namely $\sim 3\,700$, since many are clustered around R\,136. Also in
this case, only objects with $\delta_2 \le 0.1$ were considered. All
stars serving as reddening probes are shown in colour in
Figure\,\ref{fig7}. 

For RC stars, the colour excess is calculated as colour difference from
$m_{555}-m_{775}=0.95$, i.e. the nominal RC colour in these bands (see
Table\,\ref{tab1}). For UMS stars, however, we use as a reference the
isochrone shown in Figure\,\ref{fig7}, namely the zero age main sequence
(ZAMS) from the models of Marigo et al. (2008), extending up to
60\,\Msolar, and obtained specifically for the HST filters used here and
a metallicity of $Z=0.007$ as appropriate for 30\,Dor and the young LMC
population in general (e.g. Hill, Andrievsky \& Spite 1995; Geha et al.
1998). We have assumed a distance modulus $(m-M)_0 = 18.55$ (Panagia et
al. 1991; Panagia 2005; Walborn \& Blades 1997) and have already
included the intervening Galactic extinction along the line of sight,
i.e. $E(B-V)=0.07$ or $A_V=0.22$ as indicated above. Each UMS star is
then translated back to the ZAMS, along the direction of the reddening
vector, and the colour excess is computed. For stars that, once
translated to the isochrone, would be brighter than $m_{555} =13.5$ and
as such more massive than $60$\,\Msolar, we have assumed an intrinsic
colour $m_{555}-m_{775}=-0.25$.

{  Note that most of the UMS stars are likely members of binary systems,
so their intrinsic $m_{555}-m_{775}$ colour can be redder than that of
the ZAMS and our procedure could overestimate their reddening. On the
other hand, the typical mass ratio for stars above  $\sim 10$\,\Msolar\
like those in our sample is $q \simeq 0.8$ (Kiminki \& Kobulnicky 2012;
Sana et al. 2012). Model calculations of MS and PMS stars show that for
$q > 0.5$ and an age of 1\,Myr or older the $m_{555}-m_{775}$ colour
difference for these massive stars would amount to at most $0.03$\,mag
and as such it can be ignored. Also RC stars could belong to binary
systems, but since the nominal RC location is determined from the CMDs
after unsharp masking (see Figure\,\ref{fig5}), the presence of binaries
does not affect the colour excess that we derive. In fact, the excellent
match between the observed RC location in the CMD and the theoretical
models of Girardi \& Salaris (2001) indicates that possible lower-mass
companions do not appreciably alter the colour of the systems. This is
not surprising given the relatively rapid evolution of RC stars along
the red giant branch. }

\begin{figure}
\centering
%\resizebox{0.98\hsize}{!}{\includegraphics{httpfig51b.pdf}}
%\resizebox{0.98\hsize}{!}{\includegraphics{httpfig51.pdf}}
%\resizebox{0.98\hsize}{!}{\includegraphics{httpfig52.pdf}}
 \resizebox{0.98\hsize}{!}{\includegraphics{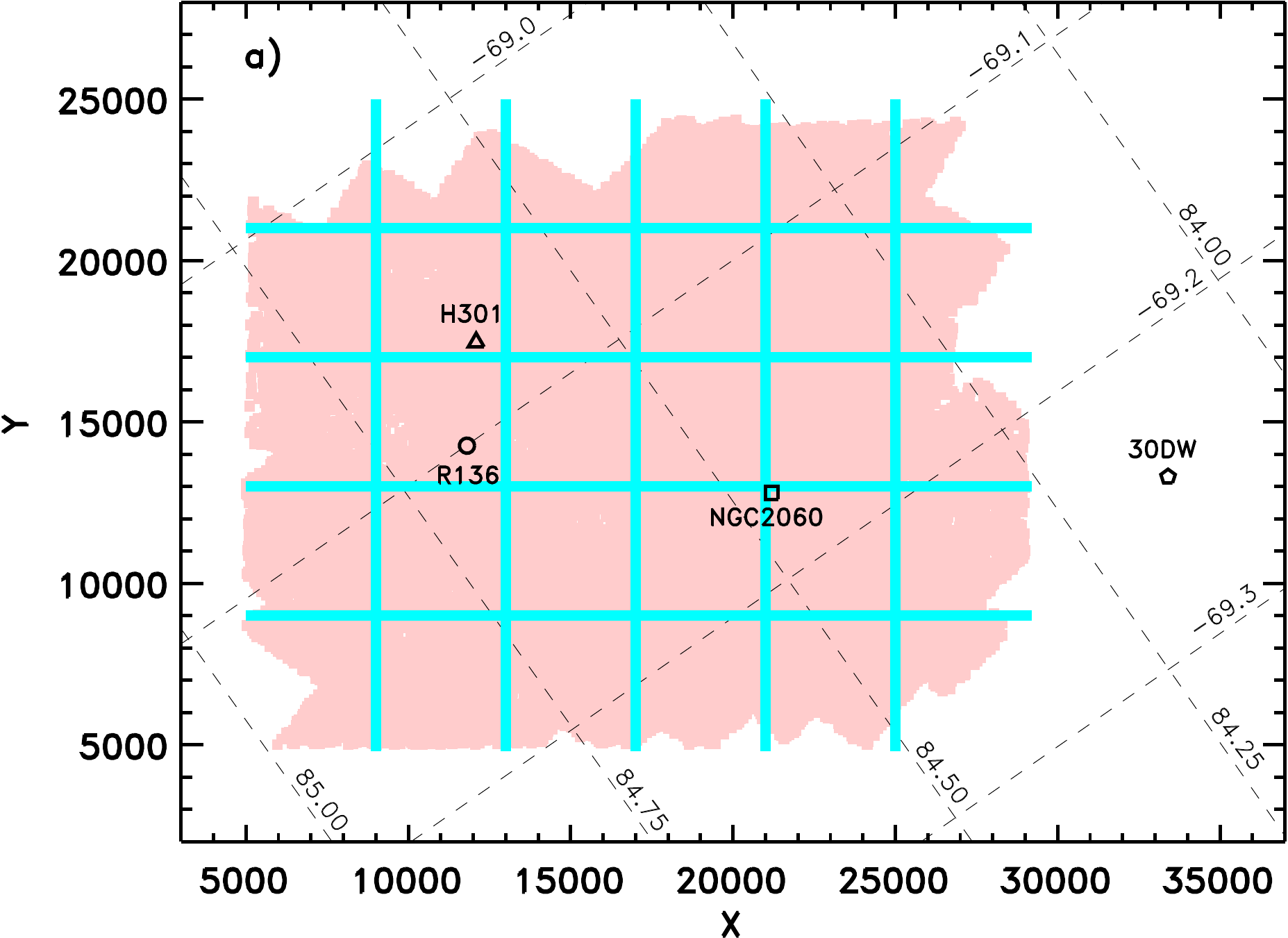}}
 \resizebox{0.98\hsize}{!}{\includegraphics{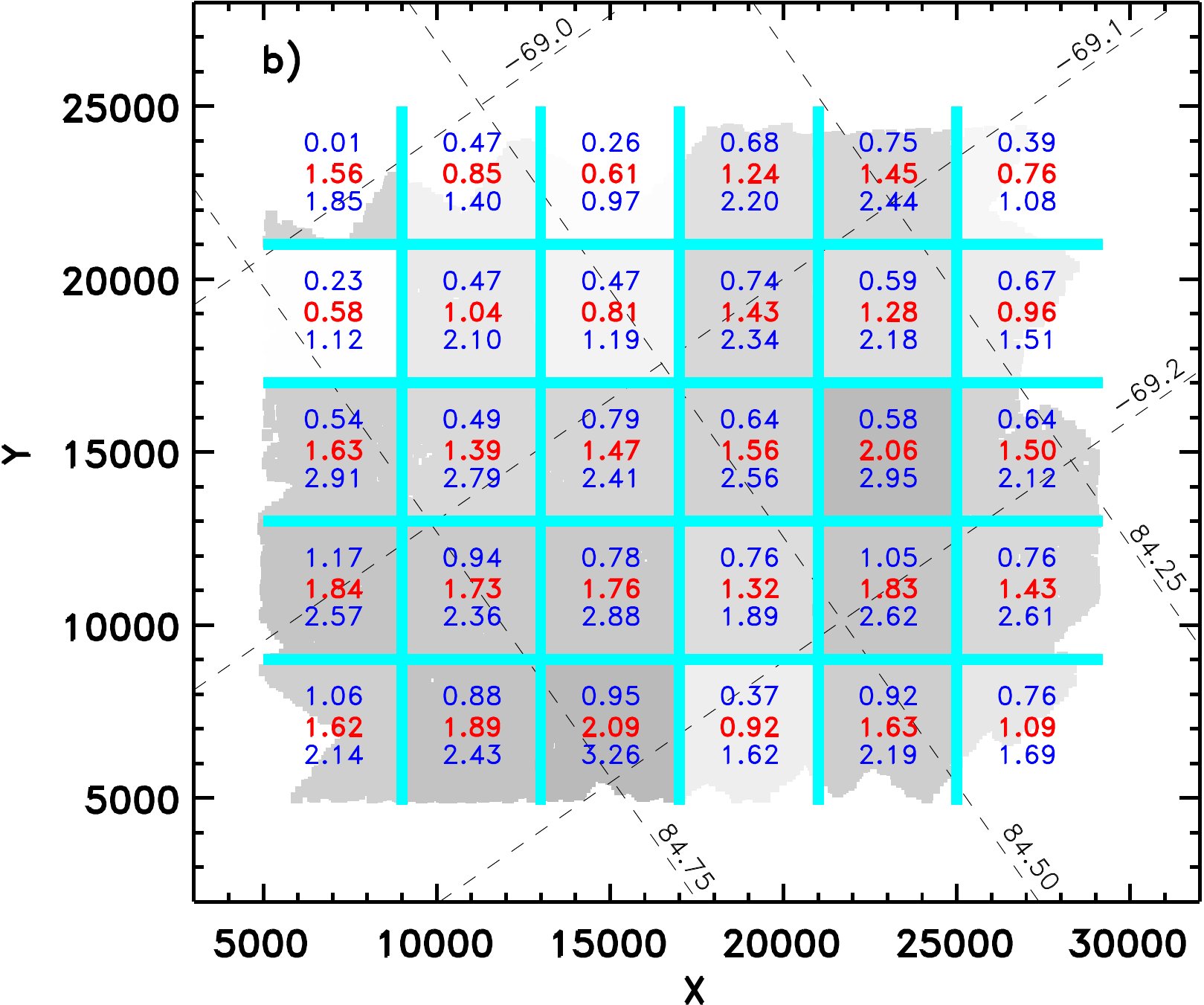}}
 \resizebox{0.98\hsize}{!}{\includegraphics{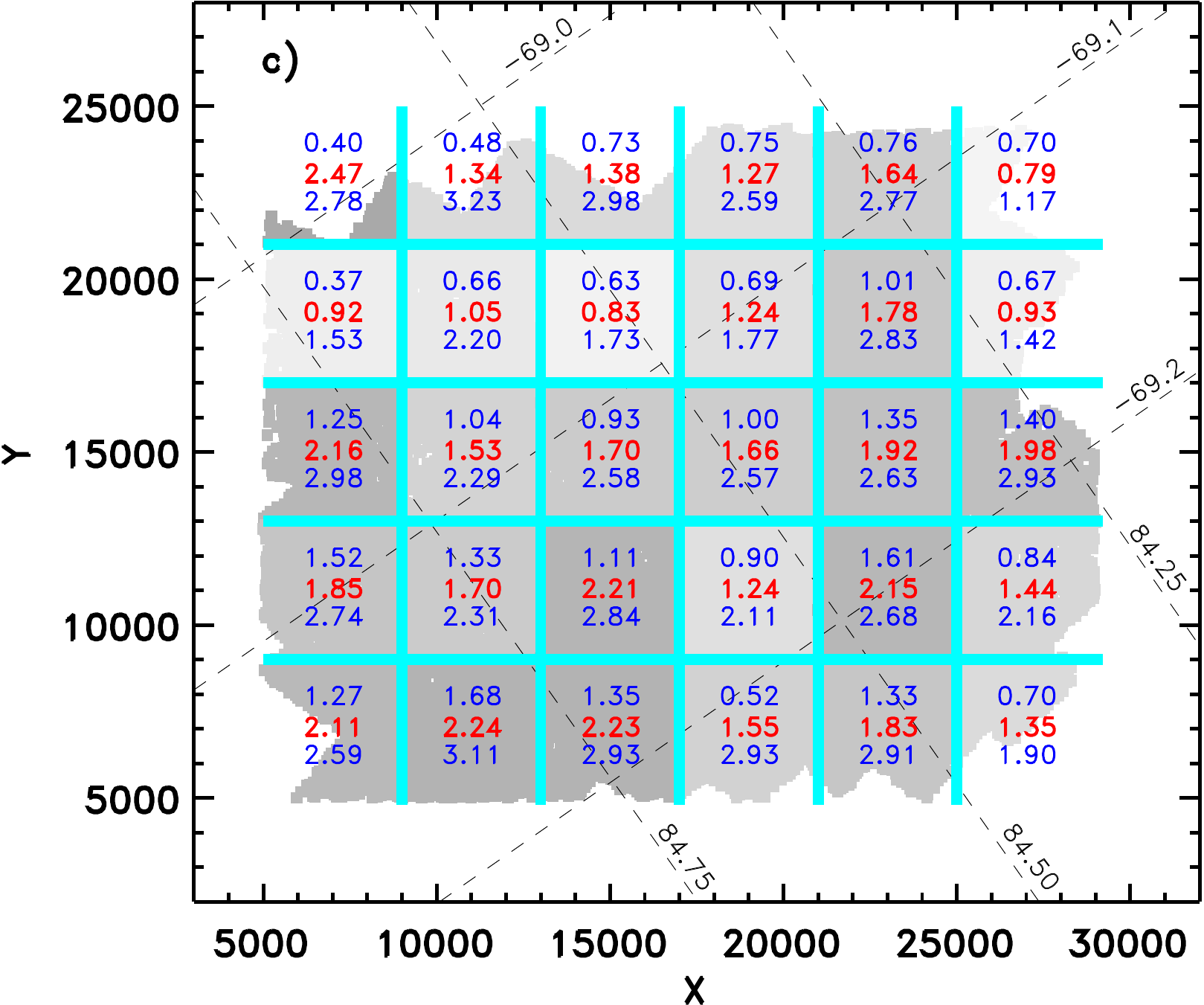}}
\caption{Panel {\em a)}: outline of the regions covered by these
observations. The centres of the cluster R\,136, Hodge 301, and
NGC\,2060 are marked and we also indicate the 30\,Dor West region 
studied by De Marchi et al. (2014). Panels {\em b)} and {\em c)}: maps
of  the reddening statistics towards RC stars and UMS stars,
respectively. The entire field  of view is divided in  cells of
$160^{\prime\prime}$ or $\sim 40$\,pc on a side. The 17, 50, and 83
percentiles of $A_{555}$ are displayed in each cell. The extent of the
shading inside each cell shows the regions covered by the observations
and the level of grey in panels {\em b)} and {\em c)} reflects the
median reddening value. The same scale of grey is used. Right ascension 
and declination are shown in decimal degrees.} \label{fig8}
\end{figure}

The resulting reddening distribution towards the selected stars is shown
schematically in Figure\,\ref{fig8}, {  panels {\em b)} and {\em c)}},
where the median and the 17 and 83 percentile values of $A_{555}$ are
displayed across the field of view in cells of $160^{\prime\prime}$ or
$\sim 40$\,pc on a side. {  For reference, in panel {\em a)} we mark
the positions of the three clusters R\,136, Hodge\,301, and NGC\,2060,
as well as the location of the 30\,Dor West field studied by De Marchi
et al. (2014).} Each cell in the panels contains on average $\sim 110$
stars of either type (RC or UMS), so the percentile values reported in
the figure are fully statistically significant, with the exception of
some cells along the borders that are only partly covered by the
observations. The extent of the shading inside each cell marks the
portion actually covered by the observations, and the level of grey
{  in panels {\em b)} and {\em c)}} reflects the median reddening. The
same scale of grey is used in both panels. 

\begin{figure}
\centering
%\resizebox{\hsize}{!}{\includegraphics[width=16cm]{httpfig58d.pdf}}
 \resizebox{\hsize}{!}{\includegraphics[width=16cm]{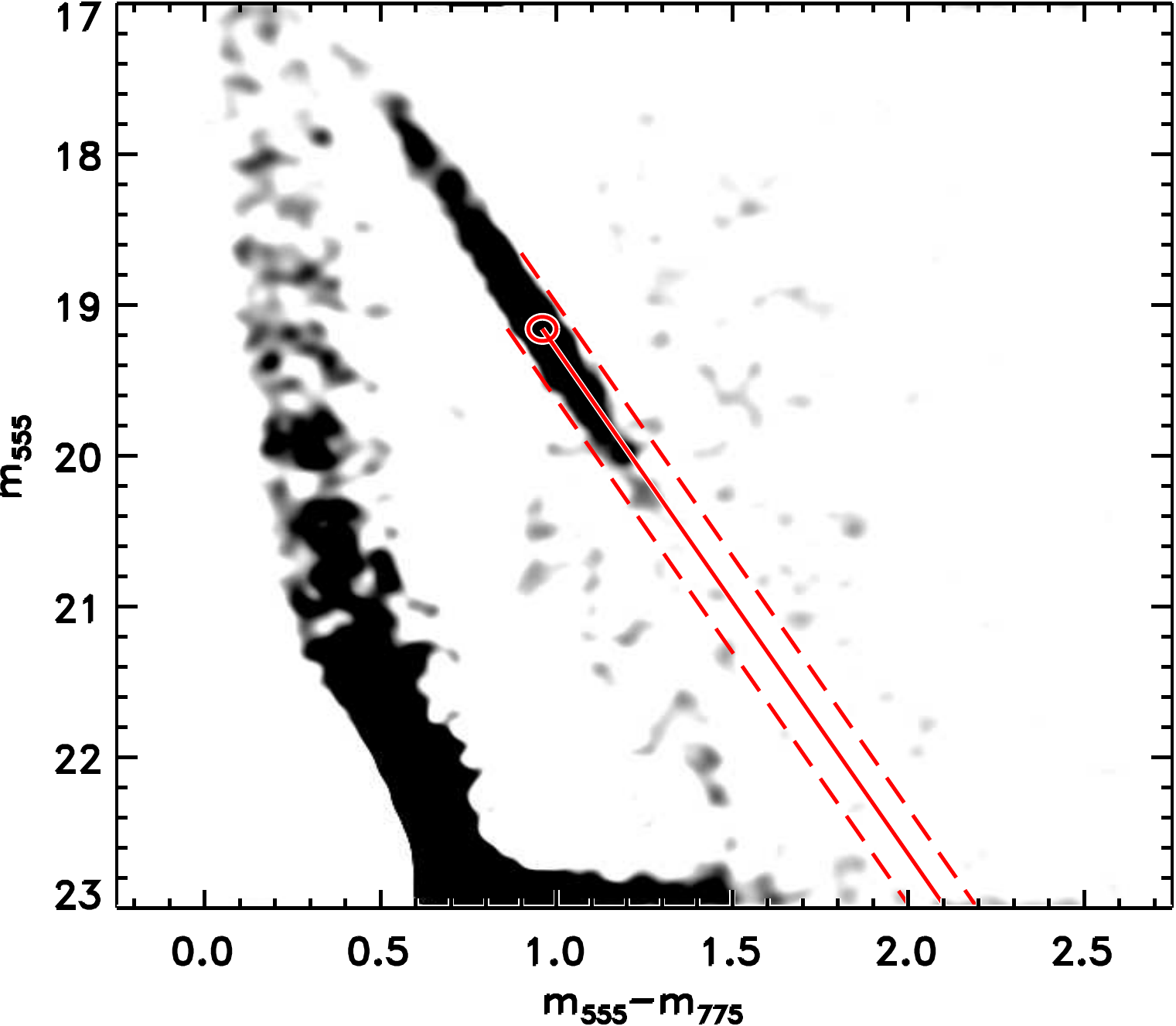}}
\caption{Same as Figure\,\ref{fig5}, but after reddening correction to 
RC stars only. We have applied to all RC stars inside each cell of 
Figure\,\ref{fig7}b the median $A_{555}$ value pertaining to that cell.
Because of the large spread of extinction values inside each cell,
applying a single reddening correction to all stars does not bring all
RC objects back to their nominal location indicated by the red ellipse.}
\label{fig9}
\end{figure}

The purpose of Figure\,\ref{fig8} is to provide a map of the typical
$A_{555}$ values and of their variations across the field, but because
of the large dispersions it should not be used as an ``extinction map''
to correct for reddening the photometry of individual objects. This
point is particularly important to understand, because the apparent
similarity of the extinction towards RC and UMS stars might be
deceiving. Indeed, the median reddening value in fully populated cells
and most of those along the borders indicate that there is comparable
extinction towards RC and UMS stars inside the same cell. It would,
therefore, be tempting to conclude that on the $\sim 40$\,pc scale of a
cell the extinction is known and that one might use Figure\,\ref{fig8}
as a ``three-dimensional reddening map'' and take the median reddening
values to correct the photometry of all stars inside the corresponding
cell. As mentioned above, this could introduce large uncertainties,
since there are systematic differences in the distribution of reddening
values between cells, even if the median values are similar

% and between UMS and RC stars inside the same cell.

{\rm As a first example, we show in Figure\,\ref{fig9} the result of
applying to all RC stars inside a cell the median $A_{555}$ value for
that cell. Note that the correction for reddening is applied on purpose
only to RC stars, not also to MS objects, so one can directly compare
Figure\,\ref{fig9} with Figure\,\ref{fig4}. It is immediately clear that
the spread of $A_{555}$ values inside each cell is so broad that using a
single value of $A_{555}$ for all the RC stars in that cell does not
bring them all back to their nominal undispersed location and the spread
remains. In  other words, without further corrections one would not be
able to derive sensible intrinsic physical parameters of RC stars from
the CMD of Figure\,\ref{fig9}, as was the case for Figure\,\ref{fig5}.}

\begin{figure}
\centering
%\resizebox{\hsize}{!}{\includegraphics[width=16cm]{httpfig56.pdf}}
 \resizebox{\hsize}{!}{\includegraphics[width=16cm]{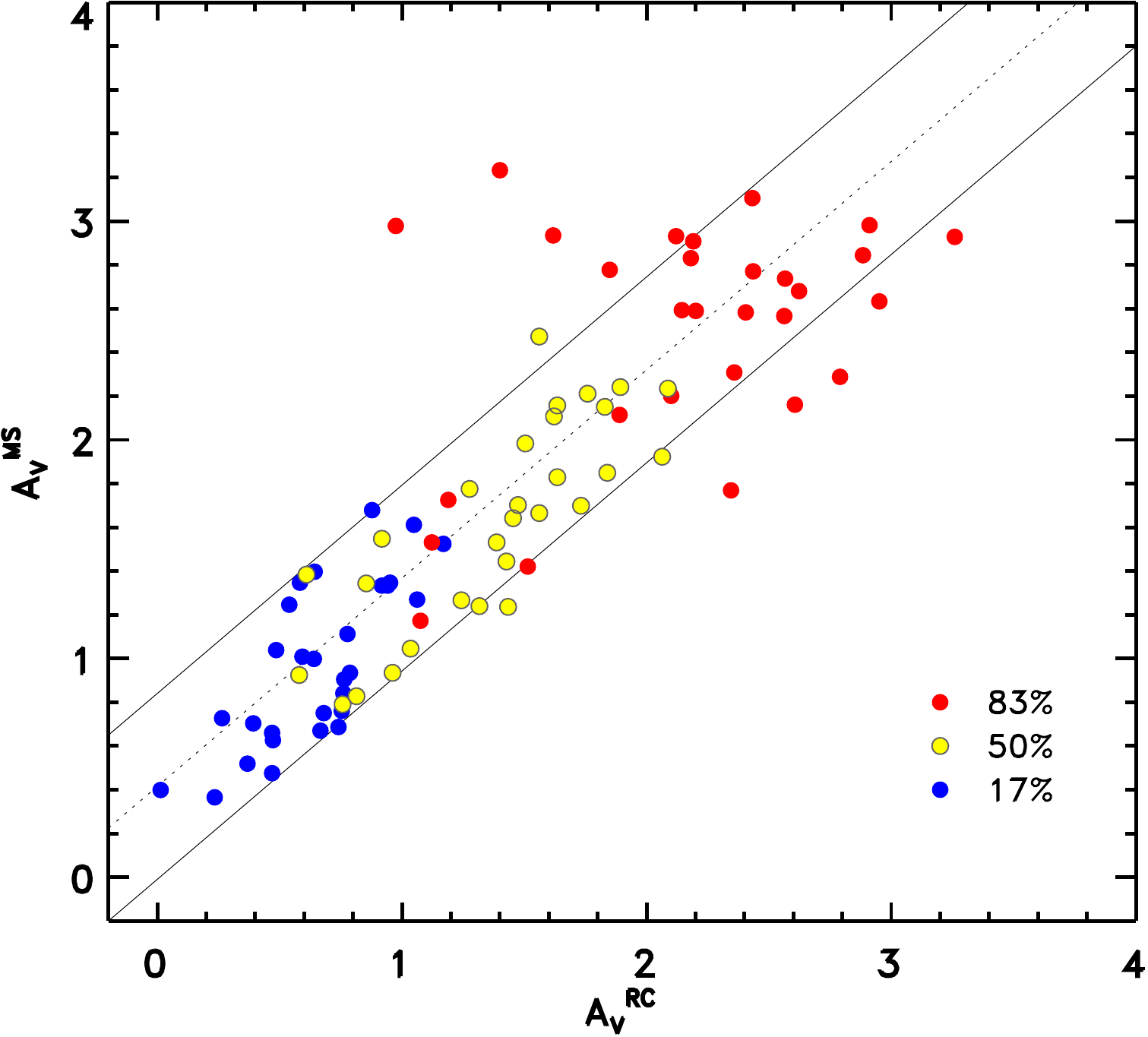}}
\caption{Relationship between reddening statistics inside the cells of
Figure\,\ref{fig8} towards RC and UMS stars. Each dot corresponds to a
cell and different colours are used for the 17, 50, and 83 percentile
levels. Within $\pm 0.4$\,mag  ($1\,\sigma$; solid lines), there is a
fair correlation between the  distributions as derived using RC and MS
stars.}
\label{fig10}
\end{figure}

{\rm Furthermore, Figure\,\ref{fig8} reveals that inside the same cell,
RC and UMS stars have very different reddening distributions and it is
not possible to use one type of stars to correct for extinction towards
the other type.} This is immediately clear from Figure\,\ref{fig10},
offering a direct comparison between the percentile reddening levels
measured towards UMS and RC stars. Different colours refer to different
levels: blue for 17\,\%, yellow for 50\,\%, and red for 83\,\%. The dots
in Figure\,\ref{fig10} correspond to the cells as marked in
Figure\,\ref{fig8}. Within $\pm 0.4$\,mag ($1\,\sigma$; solid lines),
there is a fair correlation between the reddening distribution derived
from RC and MS stars inside the same cell. This means that, in general,
inside a 40\,pc-wide cell, the amount of extinction towards RC and UMS
stars has a similar distribution, within the quoted $\pm 0.4$\,mag. It
is also clear, however, that the reddening towards RC stars begins at
systematically lower values than that towards UMS stars, by about
$0.4$\,mag. {  This is consistent with the projected spatial
distribution of RC and UMS stars observed across the region: while RC
stars are uniformly distributed, UMS stars are clearly clumped (see
Sabbi et al. 2015). This suggests that} UMS stars are distributed over a
smaller extent also along the line of sight and probe a more limited
volume of the ISM. This is not surprising, since the young UMS stars
tend to be clustered together in more compact groups, while the much
older RC stars are distributed more uniformly along the line of sight.
{  For example, in their study of the tomography of the LMC,  Haschke,
Grebel \& Duffau (2012) concluded that young populations (in that work
traced by Cepheids) have a considerably smaller extent along the line of
sight than older populations. }

Figure\,\ref{fig10} also reveals that, even inside the same cell, in
some cases UMS stars probe systematically more extinguished regions of
the ISM, as indicated by the dots in the upper left portion of the
figure. It appears that, while the 83 percentile value of the reddening 
towards RC stars in those cells is in the range $1 < A_V^{RC} < 2$, for
UMS objects it is $A_V^{MS} \simeq 3$. A similar conclusion had already
been reached by Zaritsky (1999), whose analysis of the reddening
distribution towards hot ($T_{\rm eff} > 12\,000$\,K) and cold
(5\,500\,K $< T_{\rm eff} <$ 6\,500\,K) stars contained in the
Magellanic Clouds Photometric Survey (Harris, Zaritsky \& Thompson 1997)
revealed that UMS stars are on average more extinguished than red
giants. Although this is not always the case, even around the Tarantula
nebula (see De Marchi et al. 2014), in comparison to RC objects, UMS
stars are by their very nature systematically more closely associated
with the higher density ISM regions in which they formed. Therefore, a
higher reddening level is to be expected towards UMS objects,
particularly when star formation has only recently started. {  In their
study of the main body of the LMC, Zaritsky et al. (2004) concluded
indeed that, on average, hot stars appear to be preferentially located
in dusty regions. }

{\rm In summary, our observations leave no doubt that there is a
considerable amount of dust distributed between stars along the line of
sight in this young complex. It has long been known from radio and IR 
observations of Galactic star forming regions that distributions of this
type are common in the Milky Way (e.g. Panagia 1974; Natta \& Panagia
1976; and references therein). It is important to understand whether
this is unique to the local Universe or it is typical of massive star
forming regions in general, because it appears to be is in contrast with
what has been concluded by Calzetti, Kinney \& Storchi--Bergmann (1996)
for extra-galactic star forming regions. The Tarantula complex offers
the best environment to study in detail different phases and conditions
of star formation, with no ambiguity about the distance or contamination
by background sources.} 

\subsection{How to correct for extinction}

Although the general trend in Figures\,\ref{fig8} and \ref{fig10} shows
that a fair level of correlation exists between the reddening probed by
RC and UMS stars, it also reveals that the value of $A_{555}$ is subject
to wide variations over the field. The typical difference between the 17
and 83 percentile values (blue and red dots in Figure\,\ref{fig10},
respectively) for UMS stars is $\sim 2.5$\,mag, indicating typical
fluctuations of the order of $\pm 1$\,mag in $A_{555}$ ($1\,\sigma$)
towards the young stars inside any given cell. Therefore, using a single
value of $A_{555}$ to correct the photometry of all stars in any given
cell will introduce large uncertainties on the stellar parameters ($L,
T_{\rm eff}$), with important implications for the masses and ages
derived through comparison with theoretical isochrones. 

In particular, for a typical PMS star of 1\,\Msolar, an uncertainty of
$\pm 1$\,mag in $A_V$ leads to an uncertainty of a factor of about 2 on
the age and of about $1.5$ on the mass, which grows to 2 or more for
objects younger than $\sim 4$\,Myr. Such uncertainties dominate over
those caused for instance by accretion-induced variability or unresolved
binaries (e.g. Gouliermis 2012). Therefore, since one of the goals of
the HTTP (Sabbi et al. 2013) is to determine the physical properties of
the young populations in these regions (see e.g. Cignoni et al. 2015),
including the PMS stars (De Marchi, Panagia et al. 2015, in prep.), it
is essential to apply the most appropriate correction for reddening
separately on each individual object.

Thus, the simple projected two-dimensional distribution provided by the
reddening map (in fact, a higher-resolution version of
Figure\,\ref{fig8}) must be complemented by additional information on
the properties and spatial distribution of the objects, in order to
constrain the position of the stars with respect to the extinguishing
material along the line of sight. In the following, we discuss the
conditions under which it is possible to apply an extinction correction
to individual young objects in the CMD and how to do it. We can
distinguish several cases, as follows.

\begin{enumerate}

\item 

Massive stars of spectral types O and B (blue dots in
Figure\,\ref{fig7}) spend most of their life on the UMS (e.g., Marigo et
al. 2008), so the reddening correction can be determined for each object
individually as per the procedure described above, since the extinction
law is very robust. 

\item   

For young PMS stars that are identified through their $H\alpha$ excess
emission (De Marchi et al. 2010, 2011b) and whose spatial distribution
projected on the sky overlaps with that of UMS stars, one can reasonably
assume that they are physically co-located with the UMS objects also
along the line of sight. In this case, the extinction appropriate for
each PMS star can be derived as the average of the extinction towards a
subset of UMS stars in their vicinity. For example, this ``nearest
neighbours'' approach has been used to correct for extinction the PMS in
the core of 30\,Dor (De Marchi et al. 2011a), and in the NGC\,346 star
forming cluster in the Small Magellanic Cloud (De Marchi et al. 2011b).
In both cases, using between 5 and 20 nearest neighbours on the entire
photometric catalogue resulted in a tight UMS in the CMD. It is
reasonable to adopt this approach also for stars that share the same
photometric properties as young PMS objects without necessarily
displaying $H\alpha$ excess emission, namely objects that occupy a
similar position in the broad-band CMD and whose projected spatial
distribution follows that of UMS stars. These objects could indeed be
PMS stars not actively accreting at the time of observation, as it is
known that PMS objects show large variations in their $H\alpha$ emission
over hours or days (e.g. Fernandez et al. 1995; Smith et al. 1999;
Alencar et al. 2001). 

\item

For older ($\ga 10$\,Myr) PMS objects with $H\alpha$ excess emission but
with a spatial distribution different from that of UMS stars, it is not
safe to assume that the ISM probed by the two types of objects be
similar. The UMS stars probe the most recent episode of star formation,
associated with a denser ISM, where one would expect higher extinction
values than towards older PMS objects. While one could still use the
nearest UMS stars to estimate the amount of reddening, it would likely
be too high and after correction some stars might appear bluer than the
MS. Additional constraints need to be applied in this case. A possible
approach, followed for instance by De Marchi et al. (2011a) for the
population of $\sim 12$\,Myr old PMS stars in 30\,Dor, is to de-redden
all these objects by the same amount, chosen in such a way to guarantee
a statistically acceptable distribution of colours, i.e. that no more
than $\sim 17\,\%$ of the objects are bluer than the MS after reddening
correction. Alternatively, one could estimate the extinction towards
older PMS stars individually using as nearest neighbours the RC objects
in their vicinity. As shown above, RC stars are more uniformly
distributed and sample a less dense ISM, resulting in typically lower
median extinction values towards those lines of sight (see
Figures\,\ref{fig8} and \ref{fig10}). Needless to say, both approaches
imply a larger uncertainty on the physical parameters that one can
derive for these PMS objects, and they must be properly taken into
account in any further analysis.

\item

In all other cases, and for a more solid result in case (iii), one needs
spectral information. Using large ground-based facilities it is already
possible to obtain the spectral type of a sample of bright PMS stars in
less crowded regions in the Magellanic Clouds (e.g. Kalari et al. 2014),
providing reliable intrinsic colours and effective temperatures for at
least the most massive of these objects. In the Magellanic Clouds, it
will be possible to extend this investigation to older PMS stars of
solar mass with the {\em James Webb Space Telescope} (Gardner et al.
2006).

\end{enumerate}

% We underline again that Figure\,\ref{fig8} displays maps of the
% two-dimensional extinction distribution towards both young (UMS) and
% old (RC) stars. The primary purpose of these maps is to illustrate
% the properties of the ISM and to provide the typical extinction
% values and their variations across the Tarantula nebula. To de-redden
% the photometry of individual stars, these maps (in fact a
% higher-resolution version of the same) must be supplemented by
% additional information on the properties and spatial distribution of
% the objects, in order to select meaningful nearby extinction probes
% to use in the interpolation.

Following this approach one obtains a much more reliable extinction
correction than the one achievable using pixel-to-pixel maps of the
reddening of the gas in the star forming region, estimated  from ratios
of H recombination lines at different wavelengths (e.g. Natta \& Panagia
1984; Calzetti et al. 1996; Pasquali et al. 2011; Pang, Pasquali \&
Grebel 2011;  Zeidler et al. 2015). The reason is easily understood and
can be summarised as follows.

\begin{enumerate}

\item

A relationship between the colour excess of the gas and the observed and
theoretical flux ratio of line pairs can only be defined in the ideal
(yet unlikely) case that the dust is all in the foreground and is
homogeneously distributed (see Calzetti et al. 1996). If it is not,
Natta \& Panagia (1984) have shown that the solution is not unique and
may span large uncertainties.

\item

The extinction curve specific to the environment must be known (see
Calzetti et al. 1996). As we show in this work, it is not safe to adopt
the extinction law of the Galactic diffuse ISM in star forming regions.

\item

Extinction is the sum of absorption and scattering, which affect light
differently depending on whether the photons originate from stars or
from an extended gas cloud. While for a star both absorption and
scattering result in the loss of a photon from the beam, in a nebula 
scattering simply diffuses the photon within the nebula and only
absorption will kill it. Therefore, to interpret observed H line
intensity ratios properly, a full knowledge of the dust properties
including both absorption and scattering cross sections is required.

\item

The map of gas line ratios is valid for a specific optical depth, that
of the gas. Unless stars share a similar location along the line of
sight, using these maps for the stars is a delusive strategy and the
resulting reddening correction will be systematically wrong.

\end{enumerate}

Since the above conditions are normally not satisfied in massive
clusters or regions of extended star formation (and surely not in
galaxies; see, e.g., Penner et al. 2015), it is somewhat na{\"\i}ve and
exceedingly simplistic to rely on maps of the gas line ratios for a
quantitative measure of the reddening towards individual stars. The very
fact that the extinction towards the gas is often found to be larger
than that of the UMS stars indicates that the thickness of the dust
layers in front of stars are not the same as the ones attenuating the
gas. Elementary logic dictates that loose matter located behind a source
cannot produce reliable, if any, information about what lies in front of
that source.  

Instead, the approach based on the {\em nearest neighbours with similar
age and spatial distribution} that we describe here provides what is
presently the most robust quantitative estimate of the reddening towards
individual stars in the Tarantula. Eventually, an even more accurate
reddening value for individual stars in the HTTP catalogue is expected
to be obtained through a Bayesian study of the spectral energy
distribution of {  individual objects} (Arab, Gordon, et al. 2015, in
preparation), {  including stars outside the UMS and RC loci.}

\section{Summary and conclusions}

We have studied the properties of the interstellar extinction over a
field of $16^\prime \times 13^\prime$ ($\sim 240 \times 190$\,pc$^2$) in
the Tarantula nebula, imaged with the {\em HST} as part of the HTTP
(Sabbi et al. 2013). The photometric catalogue contains more than
820\,000 stars observed at NUV, optical and NIR wavelengths through the
filters F275W, F336W, F555W, F658N, F775W,  F110W, and F160W (Sabbi et
al. 2015). Since in these regions the levels of extinction are
considerable and very uneven, RC stars are found to be spread across the
CMD defining a tight band. This has allowed us to accurately derive the
absolute extinction $A(\lambda)$ and the extinction law $R(\lambda)$ in
the range $\sim 0.3 - 1.6\,\muup$m, from more than 3\,500 RC stars. The
main results of this work can be summarised as follows.

\begin{enumerate}

\item  

The CMDs obtained from the observations reveal a prominent elongated
sequence, almost parallel to the MS, made up of several thousand RC
stars affected by various amounts of extinction (Figure\,\ref{fig2}).
%We remove from the RC sequence $\sim 1\,\%$ of the objects with
%$W_{\rm %eq}(H\alpha) > 3$\,\AA, since they could be PMS stars. 
Application of the  unsharp-masking kernel to the CMDs reduces the
contrast of the low-frequency component, resulting in a vastly improved
definition of the sharp, elongated RC feature (Figures\,\ref{fig4} and
\ref{fig5}). 

\item

From the best linear fit to the elongated RC, we obtain a fully
empirical determination of the slopes of the reddening vector in all
combinations of bands. The reddening vector appears to have a similar
slope over the entire field of view, within the uncertainty, although the
SE quadrant of the nebula reveals a systematically lower value of
$R(\lambda)$ by about 12\,\%. The excellent match between the head of
the elongated RC sequence and the position of the un-extinguished RC
predicted by theoretical models (Girardi \& Salaris 2001) is an
independent validation of the models. 

\item

The reddening slopes immediately provide the ratio $R(\lambda)$ of
total-to-selective extinction in the specific {\em HST} bands, with
high accuracy. Knowledge of the un-extinguished position of the RC in the
CMDs readily gives the absolute extinction $A(\lambda)$ in all bands
towards more than 3\,500 stars. Interpolation at the wavelengths of the
standard $B$, $V$, and $I$ bands provides the extinction curves in the
canonical forms $R_{BV}(\lambda) \equiv A(\lambda) / E(B-V)$ or
$R_{VI}(\lambda) \equiv A(\lambda) / E(V-I)$ in the range $\sim 0.3 -
1.6\,\muup$m. The latter form is more accurate because our photometry
includes observations in bands very close to the Johnson--Cousin $V$ and
$I$ filters. 

\item

{  The slope of the reddening vector in the Tarantula nebula is
considerably steeper, in all bands, }
than in the Galactic diffuse ISM, i.e. the value of $R$ is
systematically higher in 30\,Dor (Figure\,\ref{fig6}) than in the MW. We
measure $R_{BV}(V)=4.48 \pm 0.24$ and $R_{VI}(V)=3.09 \pm 0.15$ instead
of the canonical $R_{BV}(V)=3.1$ and $R_{VI}(V)=2.3$ found in the
Galaxy (e.g. Cardelli et al. 1989; Fitzpatrick \& Massa 1990).  
%%% Nor are our $R$ values or extinction law in agreement with those 
%%% measured spectroscopically by Gordon et al. (2003) towards eight 
%%% lines of sight sampling diffuse regions of the LMC outside the 
%%% Tarantula nebula and revealing systematically lower values of 
%%% $R$ [$R_{BV}(V)=2.76 \pm 0.09$]. 
On the other hand, our $R$ values are in excellent agreement with those
measured in the central NGC\,2070 cluster by De Marchi \& Panagia (2014)
from {\em HST} photometry of RC stars, that is $R_{BV}(V)=4.48 \pm
0.17$, and by Ma\'{\i}z Apell\'aniz et al. (2014) from
spectro-photometry of OB-type objects in the same field, namely,
$R_{BV}(V)=4.4 \pm 0.4$. 

\item

An immediate implication of our extinction law is that the masses
derived until now from the photometry of UMS objects have been
systematically underestimated, by a factor of $\sim 1.5$, on average,
and by more than a factor of 2 for the most extinguished 10\,\% of the
stars. For instance, the luminosity of R\,136c grows from $\log L=6.75$
to $\log L=6.9$, which according to the models of Crowther et al. (2010)
brings the star from 220\,\Msolar\ to more than 300\,\Msolar. {\rm  If
the extinction law that we measure in the Tarantula nebula is typical of
massive star forming regions in galaxies, current star formation rates
of galaxies derived from diagnostics of HII regions will have to be
seriously revised upwards.}

\item

At optical wavelengths, the extinction law $R_{VI}(\lambda)$ is best
represented by the Galactic curve shifted vertically by an offset of
$0.8$. For $\lambda > 1\,\muup$m, the best match is the Galactic law
multiplied by a factor of 2 { (both curves fall off with wavelength
as $\lambda^{-1.7}$)}. We interpret this as indication that the
Tarantula extinction curve is due to dust similar to that of the diffuse
ISM in the Galaxy, but that it contains a larger fraction of large
grains (about a factor of 2). We show that this scenario is consistent
with type II supernova explosions injecting ``fresh'' large  grains into
an otherwise MW-like mix, { as recently revealed by observations of
SN\,1987A and SN\,2010jl}. {\rm UV observations, e.g. with COS  on
board the HST, are needed to verify the evolution of the population of
grains also at the small end of the size distribution.}

\item

Since these extinction properties are consistently found across the
entire Tarantula nebula but not in the more diffuse  regions in its
surroundings, they must be related to the recent intense star formation 
episodes inside the nebula itself. Assuming that type II SNe are the
source of the extra large grains, their excess should reach a peak after
$\sim  50$\,Myr (i.e. the lifetime of the least massive type II SN
progenitors), before the grains are destroyed in the environment. The
lack of an excess of large grains in the surroundings of the Tarantula
suggests that these grains are relatively easy to destroy, making ices
in the SNe ejecta their likely source. 

%% In this scenario, the ISM in star forming regions undergoes
%% important changes over time scales of $\sim 50 - 100$\,Myr, as
%% large grains are progressively created and later destroyed. 
%% Understanding how these changes proceed in galaxies is fundamental
%% for the study of young resolved stellar populations.

\item   

Knowing the slope of the reddening vector for all bands, we can measure
the total extinction towards all objects whose nominal CMD  location can
be determined unambiguously. We select 3\,700 UMS objects and 3\,500 RC
stars to derive uniform, densely populated maps ($\sim 35$ stars per
arcmin$^2$) of the extinction towards both young and old objects. Even
though there is a fair correlation between RC and UMS reddening  over
scales of $\sim 40$\,pc, reddening towards RC stars begins at
systematically $0.4$\,mag lower values {  and UMS stars have on average
$0.4$\,mag more extinction than RC stars.} Not surprisingly, this
indicates that UMS objects sample smaller volumes along the line of
sight and probe a more limited region of the ISM. 

\item

We address the use of extinction maps for reddening correction in
regions of high and variable extinction. We show that it is  not
sufficient to rely on the projected position of the objects on the sky
and that additional information, such as age and spatial distribution,
must be used to compensate for the missing knowledge of the
line-of-sight distribution of the stars under study. We warn against the
large uncertainties inherent in applying extinction corrections based on
simple line ratios of the diffuse gas. Instead, an approach based on the
nearest RC or UMS neighbours with similar age and spatial distribution
provides a more robust quantitative estimate for individual stars.

\end{enumerate}

An important conclusion that we draw from this work is that, in regions 
of intense star formation, the ISM undergoes fundamental and rapid
changes as fresh large grains are selectively injected into it by type
II supernovae and are later destroyed. This results in profoundly
different extinction properties in these areas for periods of 50 --
100\,Myr, {\rm with $R_V$ values in the range $4 - 5$ that must be taken
into account in the study of cosmological sources. Understanding how
these changes correlate with star formation and the timescale on which
they proceed in galaxies is fundamental: not only for the study of young
resolved stellar populations in nearby galaxies, but also to decipher
the properties of star formation and chemical evolution of galaxies in
the early Universe. }

\section*{Acknowledgments}

We are grateful to an anonymous referee, whose insightful comments have
helped us to improve the presentation of this work. Support for HST
programme \#\,12939 was provided by NASA to the US team members through
a grant from the Space Telescope Science Institute, which is operated by
AURA, Inc., under NASA contract NAS 5--26555. NP acknowledges partial
support by STScI--DDRF grant D0001.82435.


\begin{thebibliography}{References}

\bibitem[]{} Alencar, S., Basri, G., Hartmann, L., Calvet, N. 2005,
  A\&A, 440, 595
\bibitem[]{} Andersen M., et al., 2009, ApJ, 707, 1347

\bibitem[]{} Bosch G., Selman F., Melnick J., Terlevich, R., 2001, 
  A\&A, 380, 137

\bibitem[]{} Brandner W., Grebel E., Barb\'a; R., Walborn N., Moneti A. 
  2001, AJ, 122, 858

\bibitem[]{} Calzetti D., Kinney A., Storchi--Bergmann T., 1996, ApJ,
  458, 132

\bibitem[]{} Cardelli J., Clayton G., Mathis J., 1989, ApJ, 345, 245

\bibitem[]{} Cardelli J., Sembach K., Mathis J., 1992, AJ, 104, 1916

\bibitem[]{} Cervi\~no M., et al., 2001, A\&A, 376, 422

\bibitem[]{} Cignoni M., et al., 2010, ApJ, 712, L63

\bibitem[]{} Cignoni M., et al., 2015, ApJ, 811, 76

\bibitem[]{} Cole A., 1998, ApJ, 500, 137

\bibitem[]{} Crowther P., et al., 2010, MNRAS, 408, 731

\bibitem[]{} De Marchi G., et al., 2011a, ApJ, 739, 27

\bibitem[]{} De Marchi G., et al., 2011b, ApJ, 740, 11

\bibitem[]{} De Marchi G., Panagia N., Girardi L., 2014, MNRAS, 438, 513

\bibitem[]{} De Marchi G., Panagia N., Romaniello M., 2010, ApJ, 715, 1

\bibitem[]{} De Marchi G., Panagia N., 2014, MNRAS, 445, 93 

\bibitem[]{} Draine, B. 2009, in Cosmic Dust Near and Far, ASP Conf. Ser. 
  414, p. 453

\bibitem[]{} Draine B., Lee H., 1984, ApJ, 285, 89 

\bibitem[]{} Dressel L., 2015, Wide Field Camera 3 Instrument Handbook,
  (Baltimore: STScI)

\bibitem[]{} Dunkin S., Crawford I., 1998, MNRAS, 298, 275

\bibitem[]{} Dwek, E. 1998, ApJ, 501, 643

\bibitem[]{} Dwek, E., Scalo, J. 1980, ApJ, 239, 193

\bibitem[]{} Evans C., et al., 2011, A\&A, 530, A108

\bibitem[]{} Fernandez, M., Ortiz, E., Eiroa, C., Miranda, L.  1995,
  A\&AS, 114, 439

\bibitem[]{} Fitzpatrick E., 1998, in Ultraviolet Astrophysics Beyond 
  the IUE Final Archive, ed. W. Wamsteker, R. Gonzalez Riestra 
  (Noordwijk: ESA), 461

\bibitem[]{} Fitzpatrick E., 1999, PASP, 111, 63 

\bibitem[]{} Fitzpatrick E., Massa D., 1990, ApJS, 72, 163

\bibitem[]{} Fitzpatrick E., Savage B., 1984, ApJ, 279, 578 

\bibitem[]{} Gardner J., et al., 2006, SSRv, 123, 485

\bibitem[]{} Gao J., Jiang B., Li A., 2009, ApJ, 707, 89

\bibitem[]{} Gall C., et al., 2014, Nature, 511, 326

\bibitem[]{} Geha M., et al., 1998, AJ, 115, 1045

\bibitem[]{} Girardi L., Groenewegen M., Weiss A., Salaris M.,
  1998, MNRAS, 301, 149

\bibitem[]{} Girardi L., Salaris M., 2001, MNRAS, 323, 109 

\bibitem[]{} Gordon K., Clayton G., Misselt K., Landolt A., 
  Wolff M., 2003, ApJ, 594, 279

\bibitem[]{} Gouliermis D., 2012, SSRv, 169, 1

\bibitem[]{} Grebel E., Chu Y.-H., 2000, AJ, 119, 787

\bibitem[]{} Greenberg J. M., 1968, in Nebulae and interstellar matter, ed. 
  B. Middlehurst, L. Aller (Chicago: Univ. Chicago Press), 221

\bibitem[]{} Harris J., Zaritsky D., Thompson I., 1997, AJ, 114, 1933 

\bibitem[]{} Haschke R., Grebel E., Duffau S., 2011, AJ, 141, 158

\bibitem[]{} Haschke R., Grebel E., Duffau S., 2012, AJ, 144, 106 

\bibitem[]{} Hess R., 1924, in Probleme der Astronomie. Festschrift fur 
  Hugo von Seeliger, (Berlin: Springer), 265.

\bibitem[]{} Hill V., Andrievsky S., Spite M., 1995, A\&A, 293, 347

\bibitem[]{} Hunter D., et al., 1995a, ApJ, 444, 758

\bibitem[]{} Hunter D., et al., 1995b, ApJ, 446, 179

\bibitem[]{} Indebetouw R., et al., 2013, ApJ, 774, 73

\bibitem[]{} Indebetouw R., et al., 2014, ApJ, 782, L2 

\bibitem[]{} Johnson H., 1968, in  Nebulae and interstellar matter, ed. 
  B. Middlehurst, L. Aller (Chicago: Univ. Chicago Press), 167

\bibitem[]{} Jones B., 1971, ApJ, 171, L57

\bibitem[]{} Kalari V., et al., A\&A, 564, L7

\bibitem[]{} Kennicutt R., 1998, ARA\&A, 36, 189

\bibitem[]{} Kiminki D., Kobulnicky H., 2012. ApJ, 751, 4

\bibitem[]{} Kylander K., Kylander O.. 1999, GIMP: The official handbook
(Scottsdale: The Coriolis Group)

\bibitem[]{} Li A., 2009, in Small Bodies in Planetary Sciences, Vol. 758, 
  ed. I. Mann, A. Nakamura, T. Mukai (Berlin: Springer), 167

\bibitem[]{} Ma\'{\i}z Apell\'aniz J., 2004, PASP, 116, 859

\bibitem[]{} Ma\'{\i}z Apell\'aniz J., et al., 2014, A\&A, 564, A63 

\bibitem[]{} Marigo P., et al., 2008, A\&A, 482, 883 

\bibitem[]{} Massa D., Savage B., Fitzpatrick E., 1983, ApJ, 266, 662

\bibitem[]{} Mathis J., Rumpl W., Nordsieck K., 1977, ApJ, 217, 425 

\bibitem[]{} Matsuura M., et al., 2011, Sci, 333, 1258

\bibitem[]{} Matteucci F., 2012, Chemical evolution of Galaxies,
(Berlin: Springer)

\bibitem[]{} Mignani R., et al., 2005, A\&431, 659

\bibitem[]{} Natta A., Panagia N., 1976, A\&A, 50, 191

\bibitem[]{} Natta A., Panagia N., 1984, ApJ, 287, 228

\bibitem[]{} Newton J., Puckett T., 2010, Central Bureau Electronic 
  Telegrams, 2532, 1

\bibitem[]{} Paczynski B., Stanek K., 1998, ApJ, 494, L219 

\bibitem[]{} Panagia N., 1974, ApJ, 192, 221

\bibitem[]{} Panagia N., 2005, in Cosmic Explosions, On the 10$^{\rm th}$
  Anniversary of SN\,1993J, IAU Coll. 192, ed. J. Marcaide, K. Weiler 
  (Berlin: Springer), 585 

\bibitem[]{} Panagia N., Gilmozzi R., Macchetto F., Adorf H.-M., 
  Kirshner R., 1991, ApJ, 380, L23 

\bibitem[]{} Pang X., Pasquali A., Grebel E., 2011, AJ, 142, 132 

\bibitem[]{} Pasquali A., et al., 2011, AJ, 141, 132

\bibitem[]{} Penner K., et al., 2015, ApJ, submitted (arXiv:1507.0728)

\bibitem[]{} Sana H., et al., 2012, Science, 337, 444

\bibitem[]{} Sabbi E., et al., 2013, AJ, 146, 53

\bibitem[]{} Sabbi E., et al., 2016, ApJS, 222, 11

\bibitem[]{} Salaris M., Girardi L., 2002, MNRAS, 337, 332

\bibitem[]{} Savage B., Sembach K., 1996, ARA\&A, 34, 279 

\bibitem[]{} Sirianni M., Nota A., Leitherer C., De Marchi G., Clampin M. 
  2000, ApJ, 533, 203

\bibitem[]{} Skorzynski W., Strobel A., Galazutdinov G., A\&A, 408, 297

\bibitem[]{} Smith, K., Lewis, G., Bonnell, I., Bunclark, P., Emerson, J.
  1999, MNRAS, 304, 367

\bibitem[]{} Spiegler G., Juris K., 1931, Phot. Korr., 67, 4

\bibitem[]{} Strom K., Strom S., Yost J., 1971, ApJ, 165, 479

\bibitem[]{} Tatton B., et al., 2013, A\&A, 554, A33

\bibitem[]{} Tognelli E., Degl'Innocenti S., Prada Moroni P. G., 2012,
  A\&A, 548, A41

\bibitem[]{} van de Hulst H., 1957, Light scattering by small particles,
  (New York: Wiley \& Sons)

\bibitem[]{} Walborn N., 1991, in Massive Stars in Starbursts, Eds. C.
  Leitherer, et al. (Cambridge: Cambridge Univ. Press), 145

\bibitem[]{} Walborn N., et al., 1999, AJ, 117, 225

\bibitem[]{} Walborn N., Blades J., 1997, ApJS, 112, 457 

\bibitem[]{} Wang S., Gao J., Jiang B., Li A., Chen, Y., 2013, ApJ, 773,
  30

\bibitem[]{} Wang S., Li A., Jiang B., 2015a, ApJ, 811, 38

\bibitem[]{} Wang S., Li A., Jiang B., 2015b, MNRAS, 454, 569

\bibitem[]{} Wang Q., Helfand D., 1991, ApJ, 370, 541

\bibitem[]{} White R., Basri G., 2003, ApJ, 582, 1109 

\bibitem[]{} Yeh S., Seaquist E., Matzener C.,  Pellegrini E., 2015,
  ApJ, 807, 117

\bibitem[]{} Yule J., 1944, Phot. J., 321

\bibitem[]{} Zaritsky D., 1999, AJ, 118, 2824 

\bibitem[]{} Zaritsky D., Harris J., Thompson I., Grebel E., 2004, 
  AJ, 128, 1606

\bibitem[]{} Zeidler P., et al., 2015, AJ, 150, 78 


\end{thebibliography}
\end{document}